\documentclass[smallcondensed]{svjour3}     % onecolumn (ditto)
\smartqed  % flush right qed marks, e.g. at end of proof
\usepackage{graphicx}
\usepackage{amsfonts,amssymb,amsopn,amstext,amscd}
\usepackage{enumerate}
\usepackage{float}
\usepackage{multicol}
\usepackage{color}
\usepackage{multirow}
\usepackage{wrapfig}
\usepackage{rotating}
\usepackage{enumitem}
\usepackage{bigints}
\usepackage{hyperref}
\newcommand{\N}{\mathbb{N}}
\newcommand{\R}{\mathbb{R}}

\allowdisplaybreaks

\newcounter{algorithm}
\journalname{RACSAM}

\begin{document}

\title{Approximate Bayesian Computation in controlled branching processes: the role of summary statistics\thanks{This research has been supported by the Ministerio de Educaci\'on, Cultura y Deporte (grant FPU13/03213), the Ministerio de Econom\'ia y Competitividad (grant MTM2015-70522-P), the Junta de Extremadura (grant IB16099) and the Fondo Europeo de Desarrollo Regional.}}


\titlerunning{ABC methodology in CBPs: the role of summary statistics}        % if too long for running head

\author{Miguel Gonz\'alez \and Rodrigo Mart\'inez \and Carmen Minuesa \and In\'es del Puerto}

\authorrunning{M. Gonz\'alez \and R. Mart\'inez \and C. Minuesa \and I. del Puerto} % if too long for running head

\institute{M. Gonz\'alez \at
              Department of Mathematics, Faculty of Sciences, and Instituto de Computaci\'on Cient\'ifica Avanzada, University of Extremadura. Avda. de Elvas s/n, 06006 Badajoz (Badajoz), Spain.  \\
ORCID: 0000-0001-7481-6561\\
              \email{mvelasco@unex.es}           %  \\
           \and
R. Mart\'inez \at
              Department of Mathematics, University Centre of Plasencia, and Instituto de Computaci\'on Cient\'ifica Avanzada, University of Extremadura. Avda. Virgen del Puerto, 2, 10600 Plasencia (C\'aceres) Spain.\\
ORCID: 0000-0003-1533-038X\\
              \email{rmartinez@unex.es}
            \and
C. Minuesa \at
              Department of Mathematics, Faculty of Sciences, University of Extremadura.\\
               Avda. de Elvas s/n, 06006 Badajoz (Badajoz), Spain.  \\
              Tel.: +34 924289300 86108\\
              ORCID:0000-0002-8858-3145\\
              \email{cminuesaa@unex.es}
              \and
           I. del Puerto \at
              Department of Mathematics, Faculty of Sciences, and Instituto de Computaci\'on Cient\'ifica Avanzada, University of Extremadura, 06006 Badajoz (Badajoz), Spain.  \\
               ORCID: 0000-0002-1034-2480\\
              \email{idelpuerto@unex.es}
}

\date{Received: date / Accepted: date}

\maketitle

\begin{abstract}
Controlled branching processes are stochastic growth population models in which the number of individuals with reproductive capacity in each generation is controlled through a random control function. The purpose of this work is to {examine the sequential Monte Carlo  Approximate} Bayesian Computation  method and to propose  appropriate summary statistics in the context of these processes. We show that the success of this methodology lies on this latter issue.   The accuracy of the proposed method is illustrated and compared with a ``likelihood free'' Markov chain Monte Carlo technique by means of a simulated example. Moreover we illustrate how to extend this methodology to a  controlled multitype branching process that has been applied to modelize real data belonging to the field of cell kinetics. Both examples are developed using the statistical software R.

\end{abstract}

\section{Introduction}

Controlled branching processes (CBPs) are a family of discrete-time stochastic processes which are appropriate to describe population dynamics. This model generalizes  the standard branching process - the so-called Bienaym\'e-Galton-Watson process. As in this latter process, each individual reproduces independently of the others and following the same distribution, referred as the offspring law. The novelty of the CBP lies in the presence of a mechanism establishing the number of individuals with reproductive capacity (progenitors) in each generation. Thus, the evolution of populations suffering from the existence of predators, populations of invasive species or different migratory movements can be modelled by using this branching process (see \cite{art-MDE} for further details). The nature of this control mechanism can be either deterministic or random - described in this latter case by what are referred  to as the control laws - and it gives rise to the models introduced by \cite{SZ} and \cite{Yanev-75}, respectively.

The recent monograph \cite{CBPs} provides an extensive description of the  probabilistic theory and inferential issues of CBPs. The great interest of the research on statistical procedures is due to the fact that the behaviour of these processes are determined by the parameters of the model associated with the offspring and control laws and in real situations those values are unknown. Focusing mainly on a Bayesian outlook, approximate Bayesian computation (ABC) methodology  has been used widely and successfully in many fields. A detailed summary on the fundamentals of ABC methods, the classical algorithms and recent developments can be found in  \cite{lin} or \cite{robert}. In particular, in the field of CBPs,
the precursor paper  \cite{art-ABC} tackled the estimation of the offspring distribution of a CBP with a deterministic control function by comparing  the rejection ABC algorithm with  a Markov chain Monte Carlo (MCMC) method.  The first one   was showed to be a good alternative  to the MCMC estimation, due to the reduction of the {computational time} while providing accurate enough {estimates}. More recently, connected with ABC methodology, \cite{drovandi2016} uses a particle MCMC method for solving inference
problems for a Bienaym\'e-Galton-Watson process which, recall, is a particular case of CBP with deterministic control function.
 %, in a non-parametric framework for the offspring distribution and assuming that the control laws belong to the family of power series distributions;
%For that purpose,  it is assumed a non-parametric framework for the offspring distribution and that the control laws belong to the family of power series distributions, by considering that only the sample made up by the population sizes is observed.

The purpose of this work is to elaborate further on ABC  inference on the class of CBPs with random control functions. For the first time, ABC methodologies are considered not only to estimate the posterior distribution which governs the reproduction law but also the ones that determine the random control on the population sizes. While the presence of the random mechanism enables us to model a greater variety of practical situations, its incorporation into the probability model makes more challenging to perform inference, as is shown in \cite{chap-proceedings-2016} in which we dealt with MCMC methods based on the observation of generation sizes. Consequently, the implementation of algorithms that improve the output of the general rejection scheme is required. We propose the application of the sequential Monte Carlo (SMC) ABC algorithm to estimate the posterior distribution of the parameters of interest.  Moreover, in order to be successful a new sample scheme respect to the one in \cite{chap-proceedings-2016} is needed. This must include information about the random control.  The performance of this  approach is appropriate under  a minimal set of assumptions and provided that it is feasible to sample from the model. The paper \cite{Beaumont-09} establishes that the  SMC ABC method can substantially outperform the rejection one when it is applied to a population genetics example. Finally, the output of SMC ABC algorithm is adjusted to account for the discrepancy between simulated and observed data by a local linear regression.

  Another important innovative feature of our current approach with respect to \cite{art-ABC}  is the inclusion and determination of an appropriate summary statistic in the development of ABC methodology in the framework of a more complex branching model. This allows us to reduce the impact of the problem known as ``the curse of dimensionality''. This issue arises when comparing large dimensional simulated  and observed data (the dimension is associated with the number of simulated generations); in those cases the discrepancy between the observed and simulated data increases as a result of a large number of {comparisons} between the components of the data. Therefore, it is better to find low dimensional summary statistics to be used in the comparison and which are informative enough about the parameters of interest. Whilst the ideal summary statistic should be a sufficient one, in the branching process setting it is complicated to determine it.  Alternatively, drawn from the knowledge about the  asymptotic properties of the model, we deduce a three-dimensional statistic that contains enough information to identify the parameters of interest. To evaluate the performance of the aforementioned algorithms we compare them with the output of  a ``likelihood free'' MCMC technique through a simulated example. It is highlighted that this methodology can be easily adaptable to more complicated branching families. In this sense, we illustrate how to extend it to a controlled multitype branching process that has been applied  to modelize real data belonging to the field of cell kinetics.

The rest of the paper is organized as follows. The probability model and notation, as well as some working assumptions, are described in Section \ref{sec:model}. Section \ref{sec:ABC} is devoted to the development of the ABC approach. % In Section  \ref{sec:MCMC}, we address the problem of sampling from the posterior distributions of interest by making use of the Gibbs sampler algorithm.
 Section \ref{sec:example} presents the simulated  and the real data examples to evaluate and illustrate the performance of the mentioned ABC algorithm. Some concluding remarks are provided in Section \ref{sec:conclusions}. Finally, additional information about the examples  are presented in the Supplementary Material.

\section{Probability model}\label{sec:model}

A \emph{controlled branching process  with random control functions} is a process \linebreak $\{Z_n\}_{n\in\N_0}$ defined as:
\begin{equation}\label{def:model}
Z_0=N_0,\quad Z_{n+1}=\sum_{j=1}^{\phi_n(Z_{n})}X_{nj},\quad n\in\N_0,
\end{equation}
where $\N_0=\N\cup\{0\}$, $N_0\in\N$, $\{X_{nj}: n\in\N_0;j\in\N\}$ and $\{\phi_n(k):n,k\in\N_0\}$ are independent families of non-negative integer valued random variables  and {the empty sum in \eqref{def:model} is considered to be 0}. The random variables $X_{nj}$, $n\in\N_0$, $j\in\N$, are assumed to be independent and identically distributed (i.i.d.) and in terms of population dynamics they represent the number of offspring given by the $j$-th progenitor of the $n$-th generation. Intuitively, the assumption on these variables means that each individual reproduces independently of the others and according to the same probability distribution. Moreover, $\{\phi_n(k)\}_{k\in\N_0}$, $n\in\N_0$, are independent stochastic processes with equal one-dimensional probability distributions. This property means that the control mechanism works in an independent manner in each generation, and once the population size at certain generation $n$, $Z_n$, is known, the probability distribution of the number of progenitors,  denoted by $\phi_n(Z_{n})$, is independent of the generation.

The common probability distribution of the random variables $X_{nj}$ is called {the} offspring  distribution or reproduction law and is denoted by $p=\{p_k\}_{k\in\N_0}$, i.e., $p_k=P[X_{nj}=k]$, $k\in\N_0$. %Its mean and variance (assumed finite) are denoted by $m$ and $\sigma^2$, respectively, and we referred to them as {the} offspring mean and variance.
Furthermore, the probability distributions of the random variables $\phi_n(k)$,   $k\in\N_0$,  called the control laws, are denoted by $\{q_j(k)\}_{j\in\N_0}$, where $q_j(k)=P[\phi_n(k)=j]$, $k,j\in\N_0$.% and we write  $\varepsilon(k)=E[\phi_0(k)]$, and $\sigma^2(k)=Var[\phi_0(k)]$ (assumed finite too) to refer to its mean and variance, respectively.

At this point we have to fix the parameters of interest and the observable sample with the aim that these can be identifiable. We consider a CBP with both the offspring  and control laws belonging to each one-dimensional parametric families and denote the offspring and control parameters by $\theta$ and $\gamma$, respectively.  Regarding the offspring law, it is usual to consider a parametric framework (see \cite{Becker-74}, \cite{art-MDE}, \cite{art-Dposterior}, \cite{Guttorp-2014}, and \cite{mmp}) since from previous observations or experiments, some information that suggests a family of distributions for the offspring law might be available (see \cite{Holgate-Lakhani-67} for further details). For instance, prokaryotic cells usually reproduce by binary fission and hence, one can parametrise the offspring distribution by considering the parameter $\theta$ defined as the probability that a cell splits off, and consequently, $1-\theta$ is the probability that a cell dies with no offspring. Another practical example is to consider a plant with a large number of seeds, where the survival of each of them (and consequently, its appearance as a plant in the following generation) is independent of the other ones and the probability that a seed grows and becomes a new plant is equal for all of them and has a small value. In this case, the Poisson distribution seems to be the appropriate distribution for the offspring distribution. The choice of  parametric control distributions is warranted given that one has different control laws for different population sizes. Then, the problem of estimating the control parameters would seem intractable based on samples with a finite dimension unless the control process is assumed to have a stable structure   {over  time}.  In practice, this information can come from the knowledge of how the population evolves. For example, if there are predators in the environment, {a binomial distribution for the number of progenitors, where the probability parameter represents the probability of survival of an individual, would be clearly justified}.

 Respect to the observed sample, for the identifiability of the model, it is necessary to get information about the reproduction and control processes.  Several {preliminary} simulation studies led us to the conclusion that {to approximate} reasonably {well} the posterior distributions of the offspring and control parameter making use of ABC methodology,  we have to assume that at least the population sizes at each generation and the number of progenitors in (at least) the  last generation are observable. The introduction in the sample of the number of progenitors in (at least) the  last generation  is crucial to identify  the  control parameter, and to define an appropriate summary statistic which {enables us} to avoid ``the  curse of dimensionality''. Actually, this data is not so difficult to be observed because of recent advances in technology. For instance, within the cell kinetics setting in \cite{art-Dposterior} a controlled two-type branching process is proposed for modelling a real data set where not only the number of individuals and progenitors are known in all the generations, but also the entire family tree. These data set is considered in Section \ref{sec:example}. Another example arises with internet protocol data modelized in a time serie context by a full observation of an INAR(1) process. These processes can be seen as a particular case of CBPs as discussed in \cite{w2011}. Two real data sets of such situations are presented in the aforementioned paper.  Consequently, let {us} consider the sample $\widetilde{\mathcal{Z}}_n =\{Z_0,\ldots,Z_n,\phi_{n-1}(Z_{n-1})\}$.

\section{ABC methodology}\label{sec:ABC}
One of the main keys to be successful in approximating the posterior distributions of the parameters of interest by applying the ABC methodology is to be able to sample from the model without being computationally costly and under not very restrictive hypotheses. This approach can be specially useful when the likelihood function is intractable.  {Our aim is} to estimate the posterior distribution  of $(\theta, \gamma)$ upon the sample $\widetilde{\mathcal{Z}}_n$, whose density is denoted by $\pi(\theta,\gamma|\widetilde{\mathcal{Z}}_n)$,  by assuming, as pointed out above, that both the offspring distribution and control laws belong to some known one-dimensional parametric families  {with unknown parameters}, that is,
\begin{equation*}
p_k=p_k(\theta),\quad  k\in\N_0, \text{ for some }\theta\in\Theta\text{ and }\Theta\subseteq\R,
\end{equation*}
and
\begin{equation*}
{q_j}(k)={{q}_j}(k,\gamma),\quad j,k\in\N_0,\text{ for some } \gamma\in\Gamma\text{ and }\Gamma\subseteq\R.
\end{equation*}
%Notice that the uncertainty is only on the exact value of the parameters.

Under the previous assumptions, the expression for the likelihood function $f(\widetilde{\mathcal{Z}}_n\mid \theta,\gamma)$ upon the sample $\widetilde{\mathcal{Z}}_n$ is complex {and practically }intractable. Indeed, since individuals reproduce independently and the control laws are independent of the offspring distribution, for any $z_0,\ldots,z_n,\phi_{n-1}^*\in\N_0$, one obtains
\begin{align}
P[Z_0&=z_0,\ldots,Z_n=z_n,\phi_{n-1}(Z_{n-1})=\phi_{n-1}^*]=P[Z_n=z_n|\phi_{n-1}(z_{n-1})=\phi_{n-1}^*]\nonumber\\
&\phantom{=}\cdot P[\phi_{n-1}(z_{n-1})=\phi_{n-1}^*]\cdot \prod_{l=1}^{n-1} P[Z_l=z_l|Z_{l-1}=z_{l-1}]\cdot P[Z_0=z_0]\nonumber\\
&=\Bigg(\delta_0(\phi_{n-1}^*)\delta_0(z_n)+(1-\delta_0(\phi_{n-1}^*))\sum_{i_1+\ldots+i_{\phi_{n-1}^*}=z_n} p_{i_1}(\theta)\cdot\ldots\cdot p_{i_{\phi_{n-1}^*}}(\theta)\Bigg)\nonumber\\
&\phantom{=}\cdot {{q}_{\phi_{n-1}^*}}(z_{n-1}, \gamma)\cdot \prod_{l=1}^{n-1}\bigg(\sum_{j=1}^\infty \Bigg(\sum_{i_1+\ldots+i_{j}=z_l} p_{i_1}(\theta)\cdot\ldots\cdot p_{i_{j}}(\theta)\Bigg)\cdot {{q}_{j}}(z_{l-1}, \gamma)\nonumber\\
&\phantom{=}+\delta_0(z_l){q_0}(z_{l-1},\gamma)\bigg)P[Z_0=z_0],\label{eq:likeli-z-tilde-general}
\end{align}
where $\delta_0(\cdot)$ denotes the Dirac delta function at 0.

 Therefore, we describe an algorithm to obtain samples from probability laws which are ``similar'' to our target distribution  and consequently, which can be used to approximate it. For this purpose, let {us} denote the observed sample by $\widetilde{\mathcal{Z}}_n^{obs}=\{Z_0^{obs},\ldots,Z_n^{obs},\phi_{n-1}(Z_{n-1})^{obs}\}$. ABC algorithms consist in sampling a large number of data from a model depending on some parameters which are generated from a prior distribution. The key idea is to identify the parameter configurations that might lead to data which are close enough to the observed sample in the sense that we specify below, and those parameters can be considered as an approximate sample from the posterior distribution $\pi(\theta,\gamma|\widetilde{\mathcal{Z}}_n^{obs})$.

In this section, built on the algorithm proposed in \cite{Beaumont-09}, we present the sequential Monte Carlo (SMC) ABC algorithm. This method consists of a number of sequential stages. At each iteration, the distribution from which the parameters are sampled is updated using the information about the parameters that were accepted at the previous step. The simulation of the data in the branching process setting is simple. Given $\theta$ and $\gamma$, first we generate the family tree of a CBP assuming that the parametric families of the offspring and control laws are known. This latter is feasible under the knowledge of the evolution of the  population considered, as was pointed out in the introduction. From the simulated data,  we obtain the population size in each generation and number of progenitors in the last one. Those data, denoted as $\widetilde{\mathcal{Z}}_n^{sim}=\{Z_0^{sim},\ldots,Z_n^{sim},\phi_{n-1}(Z_{n-1})^{sim}\}$, are compared with the observed data $\widetilde{\mathcal{Z}}_n^{obs}$ making use of a measure $\rho(\cdot,\cdot)$  {after reducing their dimension with an appropriate  summary statistic $\mathcal{S}(\cdot)$}. Moreover, we assume that $Z_0^{sim}=Z_0^{obs}$, that is, we start all the simulated processes with $Z_0^{obs}$ individuals.

The first stage involves running the ABC rejection algorithm, and in the following steps, the basic idea is to sample the parameters from a proposal distribution $\psi(\theta,\gamma)$ instead of from the prior distribution $\pi(\theta,\gamma)$ and then, to weight the accepted parameters $(\theta^{(i)},\gamma^{(i)})$ with $\omega^{(i)}\propto \pi(\theta^{(i)},\gamma^{(i)})/\psi(\theta^{(i)},\gamma^{(i)})$. To develop the SMC ABC method, it is also needed to fix a collection of tolerance levels $\epsilon_1\geq \ldots\geq\epsilon_M>0$, where $M$ is the number of iterations. The proposal distribution $\psi_t(\cdot,\cdot)$ at the $t$-th iteration is defined by using the weighted parameters selected in the previous iteration, $(\theta_{t-1}^{(i)},\gamma_{t-1}^{(i)})$, $i=1,\ldots,N$, and an auxiliary function $q_t(\cdot|\cdot)$, $t=2,\ldots, M$. To that end, as usual, we consider a mixture distribution of the weights and multivariate normal distributions as follows:
\begin{equation*}
\psi_t(\theta,\gamma)=\frac{1}{N}\sum_{i=1}^N \omega_{t-1}^{(i)}q_t(\theta,\gamma|\theta_{t-1}^{(i)},\gamma_{t-1}^{(i)}),
\end{equation*}
with $q_t(\theta,\gamma|\theta_{t-1}^{(i)},\gamma_{t-1}^{(i)})$ denoting the density function of $N\left((\theta_{t-1}^{(i)},\gamma_{t-1}^{(i)}),\sum_{t-1}\right)$, that is, a multivariate normal distribution with vector mean equal to $(\theta_{t-1}^{(i)},\gamma_{t-1}^{(i)})$ and covariance matrix, $\sum_{t-1}$. That covariance matrix is defined as twice the weighted empirical covariance matrix of the sample obtained at the previous iteration. The selection of that covariance matrix is justified by the fact that it is the optimal choice for the scale of the proposal distribution (see \cite{Beaumont-09}  {and \cite{Filippi-Barnes-Stumpf-Cornbise-2013}} for further details). Let {us} also write $\theta_t=(\theta_t^{(1)},\ldots,\theta_t^{(N)})$, $\gamma_t=(\gamma_t^{(1)},\ldots,\gamma_t^{(N)})$, and $\omega_t=(\omega_t^{(1)},\ldots,\omega_t^{(N)})$,  for $t=1,\ldots, M$.

%\vspace{0.5cm}

More precisely, the SMC ABC algorithm for a summary statistic $\mathcal{S}(\cdot)$ and the multivariate normal distribution described above as the function $q_t(\cdot|\cdot)$ consists in:
{\ttfamily
\begin{center}
SMC ABC algorithm with summary statistic $\mathcal{S}(\cdot)$ and the multivariate Gaussian distribution as $q_t(\cdot|\cdot)$
\end{center}
\begin{enumerate}
  \item [ ] Specify a decreasing sequence of tolerance levels $\epsilon_1\geq \ldots\geq\epsilon_M>0$ for $M$ iterations.
  \item [ ] For $i=1$ to $N$, do
  \begin{enumerate}
    \item [ ] Repeat
    \begin{enumerate}
    \item [ ] Sample $(\theta',\gamma')$ from the prior $\pi(\theta,\gamma)$.
    \item [ ] Sample $\widetilde{\mathcal{Z}}_n^{sim}$ from the {underlying model with offspring parameter \linebreak$\theta'$ and control parameter $\gamma'$}.%likelihood $f(\widetilde{\mathcal{Z}}_n | \theta',\gamma')$.
    \end{enumerate}
    \item [ ] Until $\rho(\mathcal{S}(\widetilde{\mathcal{Z}}_n^{sim}),\mathcal{S}(\widetilde{\mathcal{Z}}_n^{obs}))\leq\epsilon_1$.
    \item [ ] Set $(\theta_1^{(i)},\gamma_1^{(i)})=(\theta',\gamma')$.
    \item [ ] Set $\omega_1^{(i)}=1/N$.
  \end{enumerate}
  \item [ ] End for
  \item [ ] $\mathbf{\sum_1}=2\ \text{Cov}[\theta_1,\gamma_1]$ (twice the sample covariance matrix).
  \item [ ] For $t=2$ to $M$, do
  \begin{enumerate}
  \item [ ] For $i=1$ to $N$, do
  \begin{enumerate}
    \item [ ] Repeat
    \begin{enumerate}
    \item [ ] Sample $(\theta^*,\gamma^*)$ from among $(\theta_{t-1},\gamma_{t-1})$ with probabilities \linebreak $\omega_{t-1}$.
    \item [ ] Sample $(\theta',\gamma')$ from $N\left((\theta^*,\gamma^*),\mathbf{\sum_{t-1}}\right)$.
    \item [ ] Sample $\widetilde{\mathcal{Z}}_n^{sim}$ from the {underlying model with offspring \linebreak parameter $\theta'$ and control parameter $\gamma'$}.%likelihood $f(\widetilde{\mathcal{Z}}_n | \theta',\gamma')$.
    \end{enumerate}
    \item [ ] Until $\rho(\mathcal{S}(\widetilde{\mathcal{Z}}_n^{sim}),\mathcal{S}(\widetilde{\mathcal{Z}}_n^{obs}))\leq\epsilon_t$.
    \item [ ] Set $(\theta_t^{(i)},\gamma_t^{(i)})=(\theta',\gamma')$.
    \item [ ] Set $\omega_t^{(i)}\propto\pi(\theta_t^{(i)},\gamma_t^{(i)})/ \left(\sum_{k=1}^N \omega_{t-1}^{(k)}q_{t}(\theta_t^{(i)},\gamma_t^{(i)}|\theta_{t-1}^{(k)},\gamma_{t-1}^{(k)})\right)$.
  \end{enumerate}
  \item [ ] End for
  \item [ ] $\mathbf{\sum_t}=2\ \text{Cov}[\theta_t,\gamma_t]$ (twice the weighted empirical covariance matrix).
  \end{enumerate}
  \item [ ] End for
\end{enumerate}
}

Several functions can be proposed to measure the discrepancies between the simulated and the observed data. For such a purpose, we suggest the following three ``metrics'', defined for $\boldsymbol{x}=(x_1,\ldots,x_L)$, and $\boldsymbol{y}=(y_1,\ldots,y_L)\in\mathbb{R}_+^L$,  with $\mathbb{R}_+=(0,\infty)$, as:
$$ \rho_{1}(\boldsymbol{x},\boldsymbol{y}) = d_1\left(\frac{\boldsymbol{x}}{\boldsymbol{y}},\frac{\boldsymbol{y}}{\boldsymbol{x}}\right),\quad
  \rho_{e}(\boldsymbol{x},\boldsymbol{y}) = d_e\left(\frac{\boldsymbol{x}}{\boldsymbol{y}},\frac{\boldsymbol{y}}{\boldsymbol{x}}\right),\quad  \mbox{and } \quad \rho_{H}(\boldsymbol{x},\boldsymbol{y}) = d_H\left(\frac{\boldsymbol{x}}{\boldsymbol{y}},\frac{\boldsymbol{y}}{\boldsymbol{x}}\right),$$
  where $\frac{\boldsymbol{x}}{\boldsymbol{y}}=(\frac{x_1}{y_1},\ldots,\frac{x_L}{y_L})$, $\frac{\boldsymbol{y}}{\boldsymbol{x}}=(\frac{y_1}{x_1},\ldots,\frac{y_L}{x_L})$,  $d_1$ is the \emph{$\ell_1$ distance}, $d_e$ is the \emph{Euclidean distance} and $d_H$ is the \emph{Hellinger distance}. Note that the metrics $\rho_{1}$, $\rho_{e}$ and $\rho_{H}$ may not be distances in a mathematical sense, however, they keep important properties such as the non-negativity, the identity of indiscernibles and the symmetry (see \cite{art-ABC} for further metrics in the context of branching processes). It is important to highlight that any of the proposed metrics  measures the discrepancy between the components of the simulated and observed data in relative terms to avoid the influence of the different magnitudes of the coordinates. Several authors  have concluded that the choice of the distance on which the proposed metric of the ABC algorithm is based has little influence on the results, in particular in epidemic and genetic models (see \cite{mckinley},  \cite{owen} or \cite{prangle}). This finding is also supported in the context of CBPs, as we illustrate in Section \ref{sec:example}.

While the presence of a summary statistic reduces the number of comparisons in the algorithm, it also entails a lack of information from the data, unless the statistic is sufficient {(or close to sufficient)}, in which case $\pi(\theta,\gamma|\rho(\widetilde{\mathcal{Z}}_n,\widetilde{\mathcal{Z}}_n^{obs})\leq \epsilon)=\pi(\theta,\gamma|\rho(\mathcal{S}(\widetilde{\mathcal{Z}}_n),\mathcal{S}(\widetilde{\mathcal{Z}}_n^{obs}))\leq \epsilon)$. It is important to note that it is a difficult task to determine a sufficient statistic for $(\theta,\gamma)$. Now, in order to construct an appropriate summary statistic which contains relevant information about the parameters to estimate, we take advantage of the knowledge about the asymptotic  properties of the model.  Let $m$ and $\sigma^2$ denote, respectively, the  mean and variance (assumed finite) of the reproduction law, referred as offspring mean and variance. Let us write $\varepsilon(k)=E[\phi_0(k)]$, and $\sigma^2(k)=Var[\phi_0(k)]$ (assumed finite too) to refer to the mean and variance of the control laws. Under some regularity conditions {(see \cite{art-EM})} one has that, on $\{Z_n\to\infty\}$,
\begin{equation}\label{eq:conv-summary-stat}
\frac{\sum_ {i=1}^n Z_i}{\sum_{i=0}^{n-1} Z_i}\to\tau m\quad a.s.,\quad \text{and }\quad \frac{\phi_{n-1}(Z_{n-1})}{Z_{n-1}}\to \tau\quad a.s.,\quad  \text{as }n\to\infty,
\end{equation}
where $\tau=\lim_{k\to\infty}\varepsilon(k)k^{-1}$, whenever the limit exists. Consequently, the following summary statistic can be appropriate for our data:
\begin{equation}\label{eq:summary-stat}
\mathcal{S}(\widetilde{\mathcal{Z}}_n)=\left(\sum_{i=1}^n Z_i,\frac{\sum_{i=1}^n Z_i}{\sum_{i=0}^{n-1} Z_i},\frac{\phi_{n-1}(Z_{n-1})}{Z_{n-1}}\right).
\end{equation}
The first component in \eqref{eq:summary-stat} is the total progeny of the process and somehow, it represents the total {magnitude} of the process; thus, if $\rho(\cdot,\cdot)$ denotes any of the metrics defined above, then the first term in $\rho(\mathcal{S}(\widetilde{\mathcal{Z}}_n^{sim}),\mathcal{S}(\widetilde{\mathcal{Z}}_n^{obs}))$ measures the difference between the total progeny of the observed data and of the simulated data. Observe that since $m=m(\theta)$ and $\tau=\tau(\gamma)$,  the second and  the third components of $\mathcal{S}(\widetilde{\mathcal{Z}}_n)$ provide information on the offspring distribution and the control laws. These arguments explain the crucial role of the number of progenitors in (at least) the  last generation in the observed sample for our approach. Through the simulated example we show the necessity of each one of the three components in the proposed summary statistic. These features are of special interest in the cases when the functions $\theta\mapsto m(\theta)$ and $\gamma\mapsto \tau(\gamma)$ are homeomorphisms.

\begin{remark}\label{rem:cond-convergence}
In \cite{art-EM} some conditions for the convergence stated in \eqref{eq:conv-summary-stat} are established. Indeed, if the CBP $\{Z_n\}_{n\in\N_0}$ satisfies that:
\begin{enumerate}[label=(\alph*),ref=(\alph*)]
\item There exists $\tau=\lim_{k\to\infty} \varepsilon(k)k^{-1}<\infty$, and the sequence  $\{\sigma^2(k)k^{-1}\}_{k\in\N}$ is bounded,\label{rem:cond-convergence-a}
\item $\tau_m=\tau m >1$ and $Z_0$ is large enough such that $P[Z_n\rightarrow\infty]>0$,
\item $\{Z_n\tau_m^{-n}\}_{n\in\N_0}$ converges a.s. to a finite random variable $W$ such that \linebreak$P[W>0]>0$,
\item $\{W > 0\}=\{Z_n\to\infty\}$  a.s.,
\end{enumerate}
then \eqref{eq:conv-summary-stat} holds (see Proposition 3.5 in \cite{art-EM}).
\end{remark}

%\subsection{Post-processing correction method}\label{subsec:post-proc}

Finally, with the aim of improving the approximations without additional  sampling,  a local regression adjustment can be applied following the ideas in   \cite{Beaumont-02}.  The reader is referred to \cite{art-ABC} for  details in the context of CBPs with deterministic  control function. A straightforward adjustment leads to the algorithm in the case of random control functions.

\section{Examples}\label{sec:example}
\subsection{Simulated example}
\begin{wrapfigure}[15]{r}{0.38\textwidth}
\centering
\vspace*{-4ex}
\includegraphics[width=0.95\linewidth]{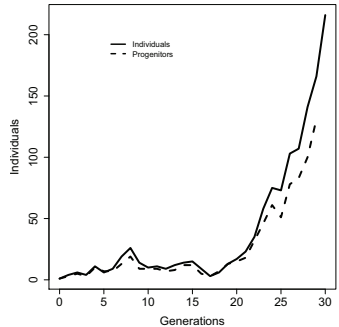}
\caption{Temporal evolution of the number of individuals (solid line) and progenitors (dashed line).}\label{im:evol}
\end{wrapfigure}
In this subsection, we present an example to illustrate the methodology proposed. We considered a CBP starting with $Z_0=1$ individual with a geometric distribution with parameter $q=0.4$ as  {the} offspring distribution and control variables $\phi_n(k)$ following a binomial distribution with parameters $\xi(k)$ and $\gamma=0.75$, where $\xi(k)= k+\lfloor\log (k)\rfloor$, for each $k\in\N$, $\xi(0)=0$, and $\lfloor x \rfloor$ denoting the integer part of a number $x$. Note that the geometric distribution arises naturally as an offspring distribution in the context of branching processes, for example, { when modelling data }from yeast cells (see \cite{Guttorp}, p.158), or in other fields as, for example, Physics (e.g. \cite{corral-garcia-font-2016}).

Observe that the control laws combine a deterministic control which is followed by a random control. In particular, these distributions enable {us} to model animal populations where new  {individuals} are incorporated into the population according to the function $\xi(\cdot)$, whereas the binomial distribution may describe the presence of predators, in such a way that $\gamma$ represents the probability that a progenitor survives and participates in the posterior evolution of the population. Under the above considerations, the offspring distribution and control laws belong to the power series family of distributions. The natural parameter of the geometric distribution as an element of the power series family of distributions is $\theta=1-q=0.6$. Regarding the offspring mean and variance, one has $m=\theta(1-\theta)^{-1}=1.5 $ and $\sigma^2=\theta(1-\theta)^{-2}=3.75$, the control means are $\varepsilon(k)=\gamma \xi(k)=0.75 \xi(k)$, $k\in\N_0$, and the asymptotic mean growth rate, referred as $\tau_m=\tau m$, is $\gamma \theta(1-\theta)^{-1}=1.125$ (see Remark \ref{rem:cond-convergence}, and notice that $\tau=\gamma$). We simulated the first 30 generations of such a CBP, whose temporal evolution is plotted in Figure \ref{im:evol} (see Table \ref{tab:sim-data} in  Supplementary M aterial for further details). In this context,  the results obtained by using the ABC methodology developed in Section \ref{sec:ABC}  are compared with the output of a ``likelihood free'' MCMC method, namely, the Gibbs sampler algorithm. In \cite{chap-proceedings-2016}, the Gibbs sampler algorithm for the sample made up by the population sizes was implemented in the context on CBPs by considering  a non-parametric framework for the offspring distribution and that the control laws belong to the family of power series distributions. Without too much difficulty one can also develop and implement the Gibbs sampler for a CBP by assuming a parametric framework for the offspring law and by considering the sample given by the population sizes in each generation and the number of progenitors in the last generation.

We implemented the SMC ABC algorithm with $M=3$ stages, that is, we updated the proposal distribution twice. We simulated pools of $9\cdot 10^4$, $9\cdot 10^5$, and $9\cdot 10^6$ of non-extinct CBPs at the corresponding iterations and fixed as the thresholds $\epsilon_1$, $\epsilon_2$ and $\epsilon_3$ the  quantiles of orders $0.025$, $0.0025$, and $0.00025$, respectively, of the sample of the distances of the simulated processes, obtained therefore a sample of length 2250 at each iteration. Thus, as suggested in \cite{Beaumont-02}, the tolerance level $\epsilon$ at the last step of the algorithm is the quantile $q_\delta$ of the sample of the distances for the simulated processes, taking $\delta=0.025\%$, that is, the sample quantile of order $2.5\cdot 10^{-4}$. Regarding the choice of the prior distribution we assumed that no information on the plausible values of the offspring and control parameters is available. Due to that, beta distributions with both parameters equal to 0.5, were used as prior  distributions for the offspring and control parameters in the ABC methodology and in the Gibbs sampler.

Based on these samples, we estimated the posterior density of the parameters $\theta$ and $\gamma$ by means of kernel density estimation. With the aim of presenting the graphs in a clearer way, we only plotted the estimates with the metric $\rho_1$ of each the posterior density in dashdotted lines in Figure \ref{im:joint-SMC-ABC}. Numerical results with the other metrics are showed in {Tables \ref{tab:summary-theta}, \ref{tab:summary-gamma} and \ref{tab:summary-taum} for the parameters $\theta$, $\gamma$, and $\tau_m$, respectively. In all cases, we have obtained that the estimates given by using $\rho_1$, $\rho_e$ and  $\rho_H$, are very similar. The estimated posterior densities differ slightly from the posteriors estimated by the Gibbs sampler. To obtain a more accurate estimation of the posterior density, we ran a post-processing algorithm by using  linear regression on the output of the  SMC ABC algorithm; the results are presented in Figure \ref{im:joint-SMC-ABC} as well and they indicate the goodness of the local linear regression adjustment. The joint posterior densities $\pi(\theta, \gamma\mid \widetilde{\mathcal{Z}}_{30})$ and  $\pi(m, \gamma\mid \widetilde{\mathcal{Z}}_{30})$ estimated by using the post-processing correction method on the output of the SMC ABC algorithm  are plotted in Figure \ref{im:SMC-proj} again only for the metric $\rho_1$.
\begin{figure}[H]
\centering
\includegraphics[width=0.45\linewidth]{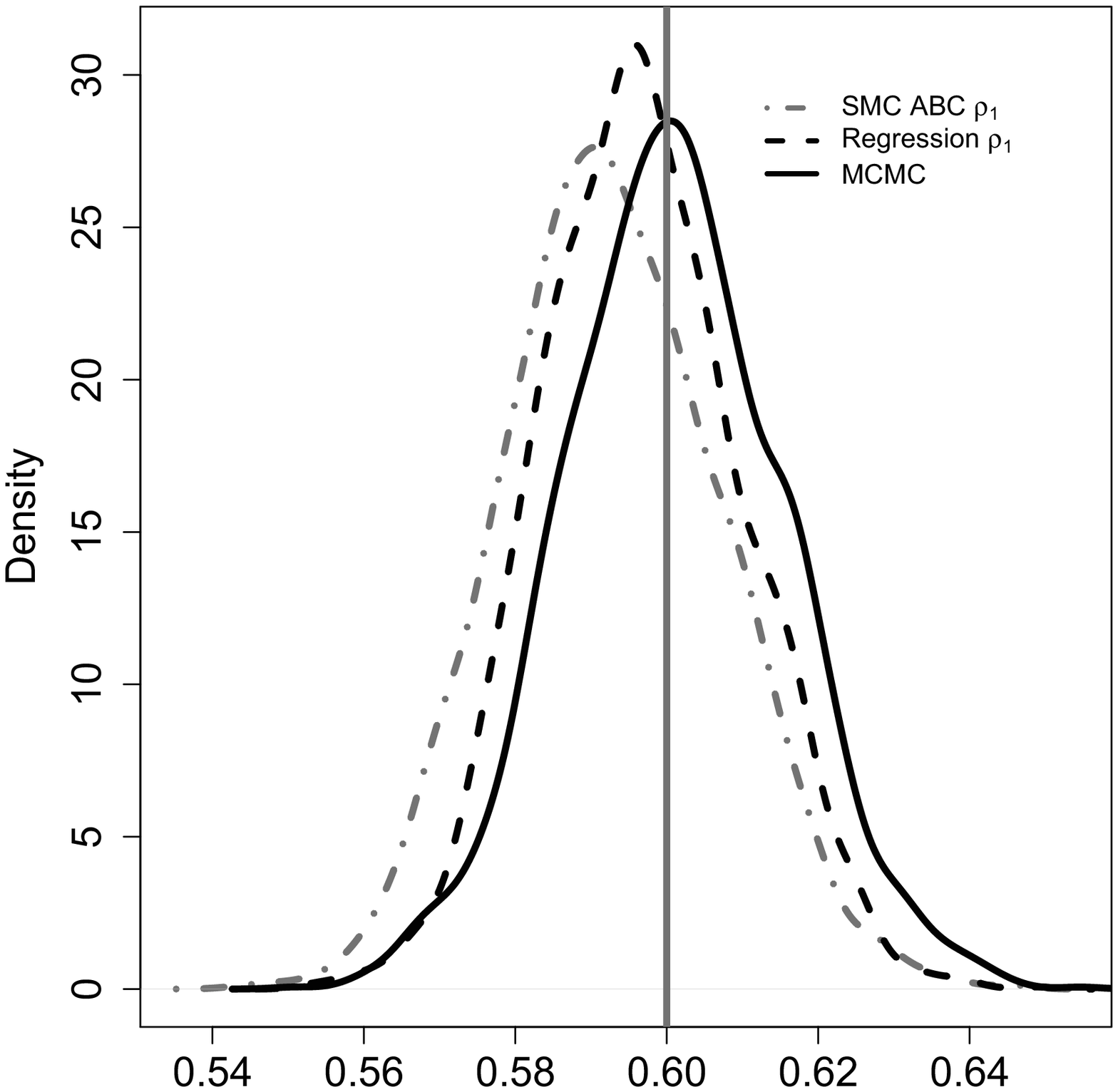}
\hspace{1cm}
\includegraphics[width=0.45\linewidth]{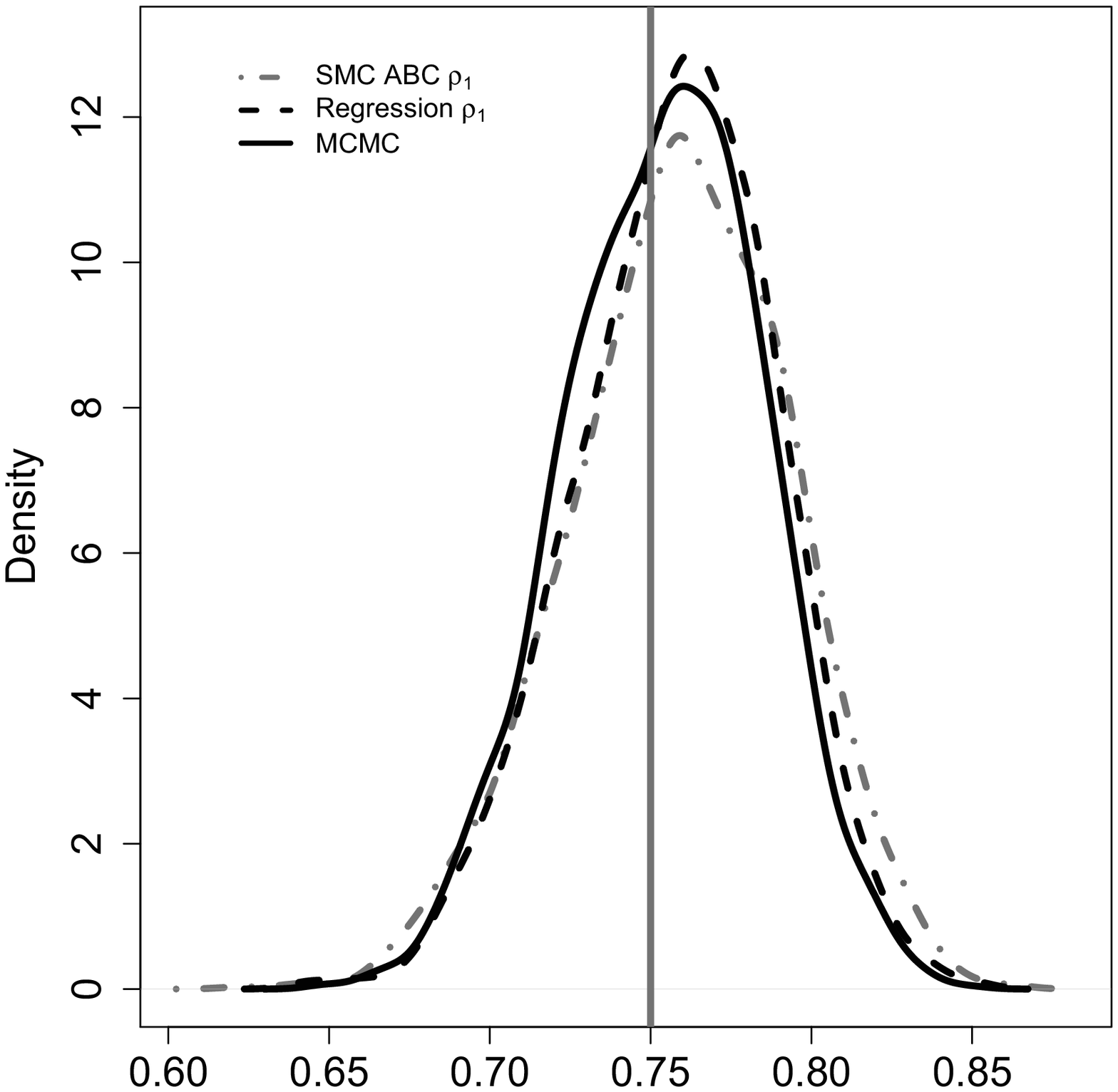}
\caption{{Posterior density functions estimated by using Gibbs sampler algorithm (solid line), the SMC ABC  algorithm (dashdotted lines) with the local linear regression adjustment (dashed lines) for the metric $\rho_1$. Left: posterior density of $\theta$. Right: posterior density of $\gamma$. Vertical lines represent the true value of the parameters.}}\label{im:joint-SMC-ABC}
\end{figure}

\begin{figure}[H]
\centering
\includegraphics[width=0.45\linewidth]{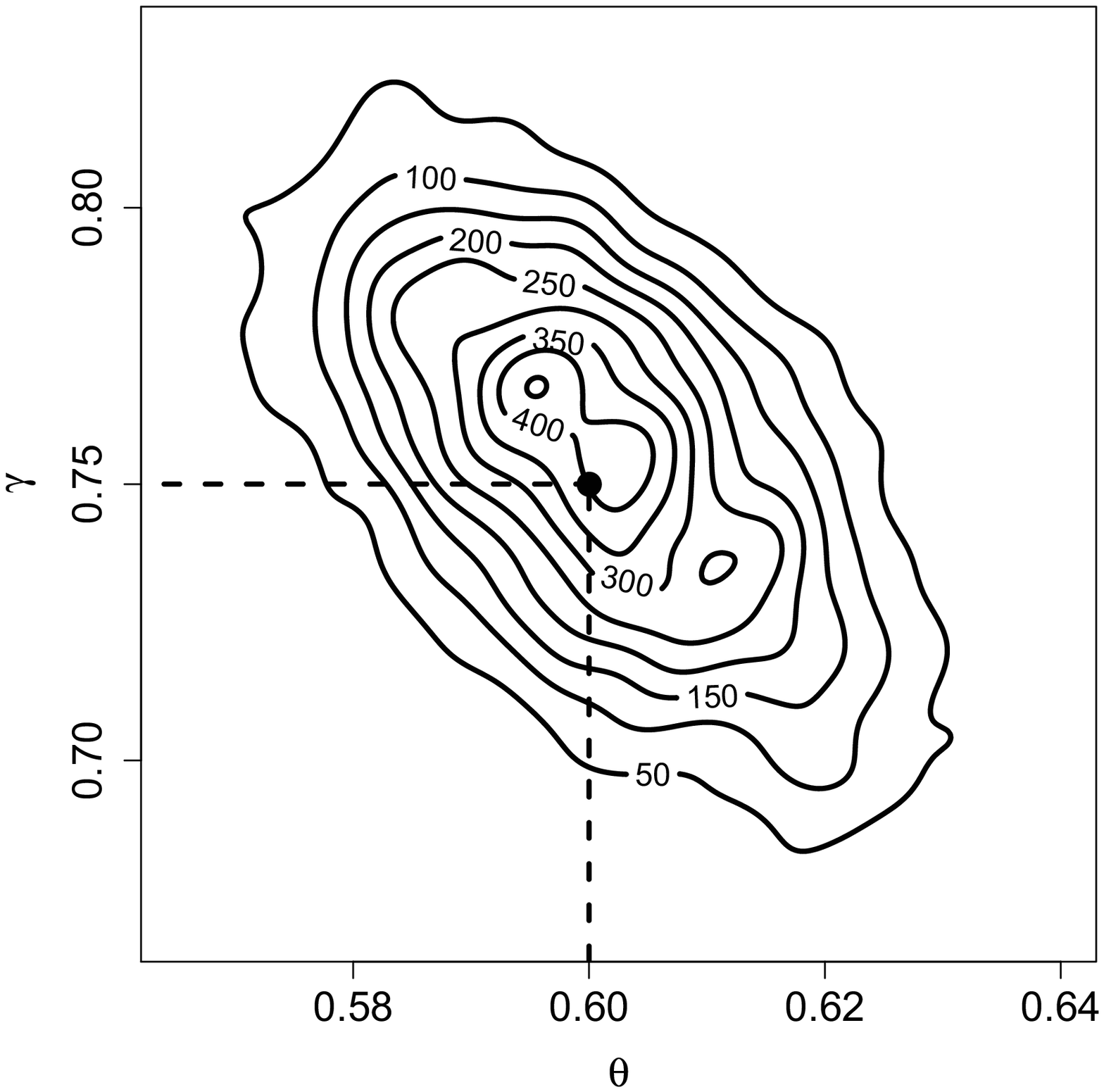}
\hspace{1cm}
\includegraphics[width=0.45\linewidth]{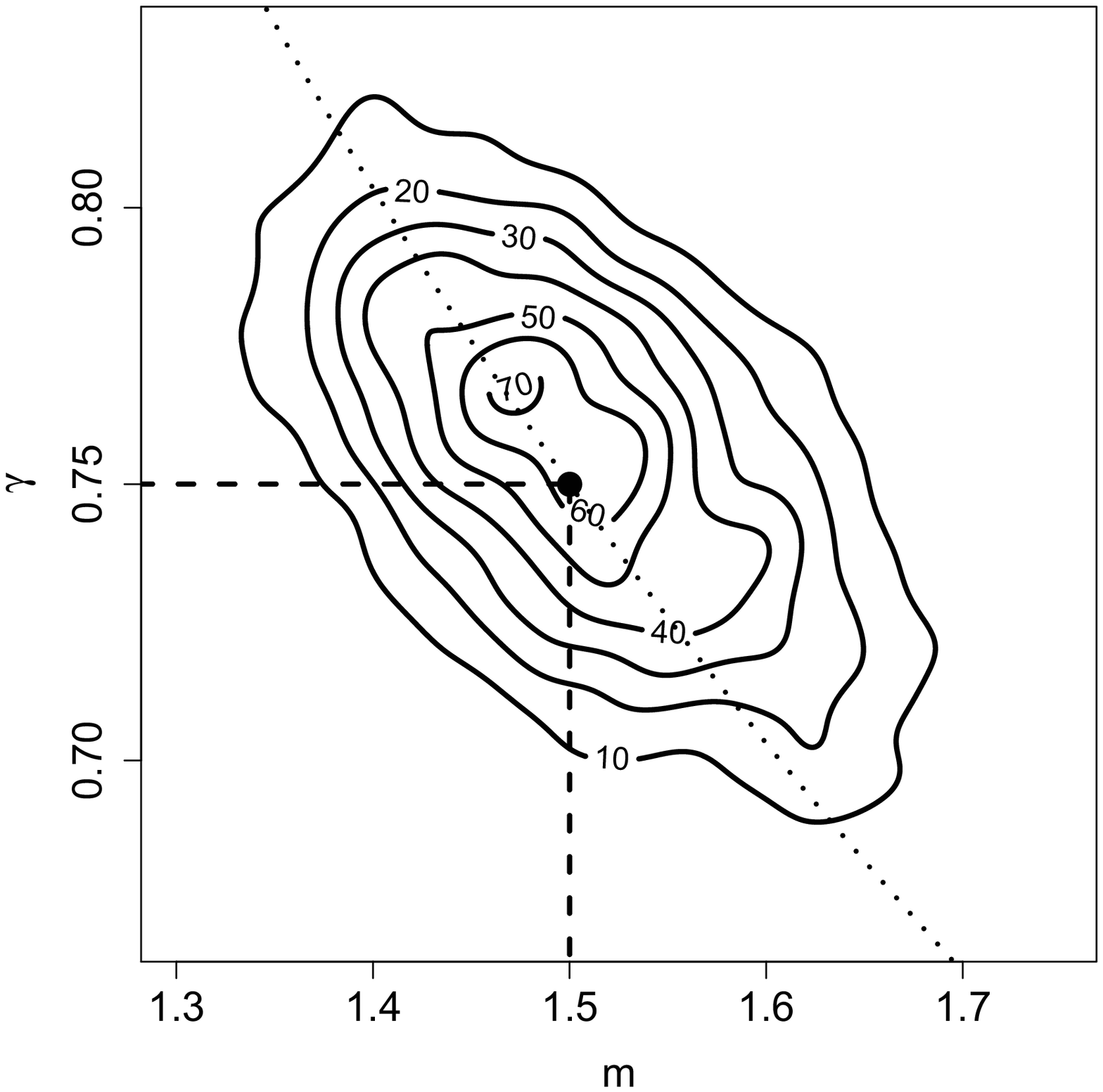}
\caption{Contour plots of the joint posterior densities estimated by the SMC ABC algorithm   with the local linear regression adjustment for the metric $\rho_1$, together with the true value of the parameters. Left: posterior density of $(\theta,\gamma)$. Right: posterior density of $(m,\gamma)$ together with the curve $\gamma m= 1.125$.}\label{im:SMC-proj}
\end{figure}
Some summary statistics to evaluate and compare the results obtained by the MCMC methodology and the ABC methods are presented in {Tables \ref{tab:summary-theta}, \ref{tab:summary-gamma} and \ref{tab:summary-taum} for the parameters $\theta$,  $\gamma$ and $\tau_m$, respectively}. We provide an estimate for the mean, variance and 95\% HPD intervals of the posterior densities based on the samples obtained by SMC ABC method and by Gibbs sampler. Moreover, we present three measures to evaluate the accuracy of the different methodologies: the relative mean square error (RMSE), the integrated squared error (ISE) and the Kullback-Leibler divergence between the posterior densities (KL). The RMSE was introduced in \cite{Beaumont-02} and it enables us to compare the mean squared error with the square of the corresponding parameter.  For the ISE and the KL,  the posterior density given by the Gibbs sampler algorithm was taken as the reference one.  In view of these comparisons, one infers that the {SMC ABC algorithm} followed by the local-linear regression adjustment {provides} the best estimate of the true posterior density functions. In addition, note that there is no significant difference among the results obtained with the different metrics.
\begin{table}
\caption{{Summary of the estimates of the posterior density $\pi(\theta|\widetilde{\mathcal{Z}}_{30})$ by the different methods.}} \label{tab:summary-theta}
\centering
\scalebox{0.9}{{\renewcommand\baselinestretch{1.5}{\small\begin{tabular}{|c|c|c|c|c|c|c|c|}
\cline{3-8}
\multicolumn{2}{c|}{} & \multicolumn{6}{|c|}{$\pi(\theta|\widetilde{\mathcal{Z}}_{30})$} \\
\cline{2-8}
\multicolumn{1}{c|}{} & Method & Mean & Variance & 95\% HPD & RMSE & ISE & KL \\
\cline{2-8}
\multicolumn{1}{c|}{} & MCMC  & $0.6009$ & $0.0002$ & $[0.5724,0.6301]$ & $0.0006$ & $\cdot$ & $\cdot$ \\
\hline
\multirow{6}{*}{\rotatebox{90}{\centering With $\mathcal{S}(\cdot)$}}
 & SMC ABC $\rho_1$             & $0.5926$ & $0.0002$ & $[0.5647,0.6203]$ & $0.0007$ & $3.1528$ & $0.1566$ \\
 & SMC ABC $\rho_e$             & $0.5933$ & $0.0002$ & $[0.5669,0.6221]$ & $0.0007$ & $2.7994$ & $0.1380$ \\
 & SMC ABC $\rho_H$             & $0.5926$ & $0.0002$ & $[0.5634,0.6207]$ & $0.0008$ & $2.9238$ & $0.1527$ \\
 & SMC ABC Regression $\rho_1$  & $0.5966$ & $0.0002$ & $[0.5718,0.6225]$ & $0.0005$ & $1.1456$ & $0.0595$ \\
 & SMC ABC Regression $\rho_e$  & $0.5965$ & $0.0002$ & $[0.5718,0.6229]$ & $0.0005$ & $1.2573$ & $0.0630$ \\
 & SMC ABC Regression $\rho_H$  & $0.5962$ & $0.0002$ & $[0.5705,0.6229]$ & $0.0005$ & $1.1171$ & $0.0623$ \\
\hline
\multirow{6}{*}{\rotatebox{90}{\centering With $\mathcal{S}_1(\cdot)$}}
 & SMC ABC $\rho_1$            & $0.6049$ & $0.00001$ & $[0.5985,0.6118]$ & $0.0001$ & $57.2624$ & $0.4455$ \\
 & SMC ABC $\rho_e$            & $0.6050$ & $0.00001$ & $[0.5989,0.6118]$ & $0.0001$ & $60.5465$ & $0.5804$ \\
 & SMC ABC $\rho_H$            & $0.6051$ & $0.00001$ & $[0.5989,0.6118]$ & $0.0001$ & $60.0393$ & $0.4972$ \\
 & SMC ABC Regression $\rho_1$ & $0.6049$ & $0.00001$ & $[0.5980,0.6118]$ & $0.0001$ & $57.1825$ & $0.4524$ \\
 & SMC ABC Regression $\rho_e$ & $0.6050$ & $0.00001$ & $[0.5989,0.6118]$ & $0.0001$ & $60.6781$ & $0.5739$ \\
 & SMC ABC Regression $\rho_H$ & $0.6051$ & $0.00001$ & $[0.5989,0.6118]$ & $0.0001$ & $60.0351$ & $0.4980$ \\
\hline
\multirow{6}{*}{\rotatebox{90}{\centering With $\mathcal{S}_2(\cdot)$}}
 & SMC ABC $\rho_1$             & $0.5746$ & $0.0003$ & $[0.5402,0.6083]$ & $0.0027$ & $17.2930$ & $1.0968$ \\
 & SMC ABC $\rho_e$             & $0.5744$ & $0.0003$ & $[0.5402,0.6096]$ & $0.0027$ & $17.1637$ & $1.0769$ \\
 & SMC ABC $\rho_H$             & $0.5745$ & $0.0003$ & $[0.5391,0.6089]$ & $0.0027$ & $16.8155$ & $1.0757$ \\
 & SMC ABC Regression $\rho_1$  & $0.5746$ & $0.0003$ & $[0.5398,0.6078]$ & $0.0027$ & $17.3233$ & $1.1028$ \\
 & SMC ABC Regression $\rho_e$  & $0.5744$ & $0.0003$ & $[0.5402,0.6096]$ & $0.0027$ & $17.1740$ & $1.0774$ \\
 & SMC ABC Regression $\rho_H$  & $0.5745$ & $0.0003$ & $[0.5396,0.6089]$ & $0.0027$ & $16.7786$ & $1.0766$ \\
\hline
\multirow{6}{*}{\rotatebox{90}{\centering With $\mathcal{S}_3(\cdot)$}}
 & SMC ABC $\rho_1$             & $0.6160$ & $0.0033$ & $[0.5164,0.7324]$ & $0.0098$ & $11.3148$ & $0.9253$ \\
 & SMC ABC $\rho_e$             & $0.6184$ & $0.0033$ & $[0.5189,0.7341]$ & $0.0102$ & $11.4357$ & $0.9422$ \\
 & SMC ABC $\rho_H$             & $0.6140$ & $0.0032$ & $[0.5171,0.7274]$ & $0.0094$ & $11.0885$ & $0.9021$ \\
 & SMC ABC Regression $\rho_1$  & $0.6182$ & $0.0032$ & $[0.5194,0.7347]$ & $0.0099$ & $11.1805$ & $0.9121$ \\
 & SMC ABC Regression $\rho_e$  & $0.6218$ & $0.0033$ & $[0.5220,0.7359]$ & $0.0104$ & $11.2876$ & $0.9282$ \\
 & SMC ABC Regression $\rho_H$  & $0.6180$ & $0.0031$ & $[0.5221,0.7307]$ & $0.0096$ & $10.8141$ & $0.8767$ \\
\hline
\end{tabular} }}}
\end{table}

\begin{table}[h!]
\caption{{Summary of the estimates of the posterior density $\pi(\gamma|\widetilde{\mathcal{Z}}_{30})$ by the different methods.}} \label{tab:summary-gamma}
 \centering \centering
\scalebox{0.9}{{\renewcommand\baselinestretch{1.5}{\small\begin{tabular}{|c|c|c|c|c|c|c|c|}
\cline{3-8}
\multicolumn{2}{c|}{} & \multicolumn{6}{|c|}{$\pi(\gamma|\widetilde{\mathcal{Z}}_{30})$}\\
\cline{2-8}
\multicolumn{1}{c|}{} & Method & Mean & Variance & 95\% HPD & RMSE & ISE & KL \\
\cline{2-8}
\multicolumn{1}{c|}{} & MCMC & $0.7539$ & $0.0009$ & $[0.6922,0.8115]$ & $0.0017$ & $\cdot$ & $\cdot$ \\
\hline
\multirow{6}{*}{\rotatebox{90}{\centering With $\mathcal{S}(\cdot)$}}
 & SMC ABC $\rho_1$             & $0.7589$ & $0.0012$ & $[0.6896,0.8248]$ & $0.0022$ & $0.2182$ & $0.0279$ \\
 & SMC ABC $\rho_e$             & $0.7593$ & $0.0012$ & $[0.6928,0.8239]$ & $0.0023$ & $0.2451$ & $0.0318$ \\
 & SMC ABC $\rho_H$             & $0.7600$ & $0.0012$ & $[0.6923,0.8270]$ & $0.0023$ & $0.2536$ & $0.0353$ \\
 & SMC ABC Regression $\rho_1$  & $0.7577$ & $0.0010$ & $[0.6945,0.8164]$ & $0.0018$ & $0.0988$ & $0.0099$ \\
 & SMC ABC Regression $\rho_e$  & $0.7584$ & $0.0010$ & $[0.6976,0.8204]$ & $0.0019$ & $0.1056$ & $0.0140$ \\
 & SMC ABC Regression $\rho_H$  & $0.7587$ & $0.0010$ & $[0.6954,0.8199]$ & $0.0019$ & $0.1542$ & $0.0176$ \\
\hline
\multirow{6}{*}{\rotatebox{90}{\centering With $\mathcal{S}_1(\cdot)$}}
 & SMC ABC $\rho_1$            & $0.7870$ & $0.0001$ & $[0.7715,0.8012]$ & $0.0025$ & $32.7316$ & $0.4832$ \\
 & SMC ABC $\rho_e$            & $0.7867$ & $0.0001$ & $[0.7710,0.8012]$ & $0.0025$ & $33.6205$ & $0.3896$ \\
 & SMC ABC $\rho_H$            & $0.7868$ & $0.0001$ & $[0.7719,0.8008]$ & $0.0025$ & $34.7825$ & $0.5141$ \\
 & SMC ABC Regression $\rho_1$ & $0.7870$ & $0.0001$ & $[0.7715,0.8012]$ & $0.0025$ & $32.7434$ & $0.4812$ \\
 & SMC ABC Regression $\rho_e$ & $0.7867$ & $0.0001$ & $[0.7710,0.8012]$ & $0.0025$ & $33.6132$ & $0.3890$ \\
 & SMC ABC Regression $\rho_H$ & $0.7868$ & $0.0001$ & $[0.7723,0.8008]$ & $0.0025$ & $34.7769$ & $0.5142$ \\
\hline
\multirow{6}{*}{\rotatebox{90}{\centering With $\mathcal{S}_2(\cdot)$}}
 & SMC ABC $\rho_1$            & $0.7483$ & $0.0019$ & $[0.6576,0.8346]$ & $0.0034$ & $0.5341$ & $0.0932$ \\
 & SMC ABC $\rho_e$            & $0.7502$ & $0.0020$ & $[0.6585,0.8359]$ & $0.0035$ & $0.5467$ & $0.0980$ \\
 & SMC ABC $\rho_H$            & $0.7498$ & $0.0020$ & $[0.6582,0.8360]$ & $0.0036$ & $0.5259$ & $0.0958$ \\
 & SMC ABC Regression $\rho_1$ & $0.7484$ & $0.0019$ & $[0.6590,0.8337]$ & $0.0034$ & $0.5198$ & $0.0918$ \\
 & SMC ABC Regression $\rho_e$ & $0.7503$ & $0.0020$ & $[0.6585,0.8355]$ & $0.0035$ & $0.5494$ & $0.0970$ \\
 & SMC ABC Regression $\rho_H$ & $0.7497$ & $0.0020$ & $[0.6569,0.8356]$ & $0.0036$ & $0.5248$ & $0.0942$ \\
 \hline
\multirow{6}{*}{\rotatebox{90}{\centering With $\mathcal{S}_3(\cdot)$}}
 & SMC ABC $\rho_1$            & $0.7116$ & $0.0251$ & $[0.4044,1]$      & $0.0472$ & $6.1637$ & $1.2377$ \\
 & SMC ABC $\rho_e$            & $0.7059$ & $0.0255$ & $[0.4064,0.9951]$ & $0.0488$ & $6.3199$ & $1.2788$ \\
 & SMC ABC $\rho_H$            & $0.7171$ & $0.0245$ & $[0.4108,1]$      & $0.0455$ & $6.0280$ & $1.2004$ \\
 & SMC ABC Regression $\rho_1$ & $0.7089$ & $0.0249$ & $[0.4039,0.9991]$ & $0.0472$ & $6.1404$ & $1.2301$ \\
 & SMC ABC Regression $\rho_e$ & $0.7001$ & $0.0253$ & $[0.3969,0.9898]$ & $0.0494$ & $6.2901$ & $1.2697$ \\
 & SMC ABC Regression $\rho_H$ & $0.7094$ & $0.0243$ & $[0.4057,0.9920]$ & $0.0460$ & $5.9933$ & $1.1895$ \\
 \hline
\end{tabular} }}}
\end{table}

\begin{table}[h!]
\caption{{Summary of the estimates of the posterior density $\pi(\tau_m|\widetilde{\mathcal{Z}}_{30})$ by the different methods.}} \label{tab:summary-taum}
\centering
\scalebox{0.9}{{\renewcommand\baselinestretch{1.5}{\small\begin{tabular}{|c|c|c|c|c|c|c|c|}
\cline{3-8}
\multicolumn{2}{c|}{} & \multicolumn{6}{|c|}{$\pi(\tau_m|\widetilde{\mathcal{Z}}_{30})$} \\
\cline{2-8}
\multicolumn{1}{c|}{} & Method & Mean & Variance & 95\% HPD & RMSE & ISE & KL \\
\cline{2-8}
\multicolumn{1}{c|}{} & MCMC  & $1.1358$ & $0.0032$ & $[1.0188,1.2418]$ & $0.0026$ & $\cdot$ & $\cdot$ \\
\hline
\multirow{6}{*}{\rotatebox{90}{\centering With  $\mathcal{S}(\cdot)$}}
 & SMC ABC $\rho_1$             & $1.1045$ & $0.0026$ & $[1.0004,1.2050]$ & $0.0024$ & $0.0048$ & $0.1760$ \\
 & SMC ABC $\rho_e$             & $1.1078$ & $0.0024$ & $[1.0093,1.2050]$ & $0.0022$ & $0.0010$ & $0.1476$ \\
 & SMC ABC $\rho_H$             & $1.1060$ & $0.0026$ & $[1.0093,1.2139]$ & $0.0023$ & $0.0039$ & $0.1576$ \\
 & SMC ABC Regression $\rho_1$  & $1.1209$ & $0.0023$ & $[1.0182,1.2139]$ & $0.0019$ & $0.0001$ & $0.0588$ \\
 & SMC ABC Regression $\rho_e$  & $1.1212$ & $0.0022$ & $[1.0271,1.2139]$ & $0.0018$ & $0.0002$ & $0.0651$ \\
 & SMC ABC Regression $\rho_H$  & $1.1208$ & $0.0024$ & $[1.0271,1.2228]$ & $0.0019$ & $0.00004$ & $0.0613$ \\
\hline
\multirow{6}{*}{\rotatebox{90}{\centering With $\mathcal{S}_1(\cdot)$}}
 & SMC ABC $\rho_1$            & $1.2049$ & $0.0002$ & $[1.1783,1.2361]$ & $0.0052$ & $0.0014$ & $0.9778$ \\
 & SMC ABC $\rho_e$            & $1.2052$ & $0.0002$ & $[1.1783,1.2361]$ & $0.0053$ & $0.0014$ & $1.0833$ \\
 & SMC ABC $\rho_H$            & $1.2055$ & $0.0002$ & $[1.1739,1.2361]$ & $0.0053$ & $0.0014$ & $1.1072$ \\
 & SMC ABC Regression $\rho_1$ & $1.2049$ & $0.0002$ & $[1.1783,1.2361]$ & $0.0052$ & $0.0014$ & $0.9798$ \\
 & SMC ABC Regression $\rho_e$ & $1.2052$ & $0.0002$ & $[1.1783,1.2361]$ & $0.0053$ & $0.0014$ & $1.0859$ \\
 & SMC ABC Regression $\rho_H$ & $1.2055$ & $0.0002$ & $[1.1739,1.2361]$ & $0.0053$ & $0.0014$ & $1.1059$ \\
\hline
\multirow{6}{*}{\rotatebox{90}{\centering With  $\mathcal{S}_2(\cdot)$}}
 & SMC ABC $\rho_1$             & $1.0108$ & $0.0029$ & $[0.9071,1.1205]$ & $0.0126$ & $2.1586$ & $1.9624$ \\
 & SMC ABC $\rho_e$             & $1.0126$ & $0.0028$ & $[0.9115,1.1161]$ & $0.0122$ & $2.1129$ & $1.9290$ \\
 & SMC ABC $\rho_H$             & $1.0122$ & $0.0029$ & $[0.9111,1.1200]$ & $0.0123$ & $2.1583$ & $1.9917$ \\
 & SMC ABC Regression $\rho_1$  & $1.0108$ & $0.0029$ & $[0.9071,1.1205]$ & $0.0126$ & $2.1489$ & $1.9611$ \\
 & SMC ABC Regression $\rho_e$  & $1.0126$ & $0.0028$ & $[0.9115,1.1161]$ & $0.0122$ & $2.1085$ & $1.9298$ \\
 & SMC ABC Regression $\rho_H$  & $1.0122$ & $0.0029$ & $[0.9067,1.1200]$ & $0.0123$ & $2.1514$ & $1.9920$ \\
 \hline
\multirow{6}{*}{\rotatebox{90}{\centering With  $\mathcal{S}_3(\cdot)$}}
 & SMC ABC $\rho_1$             & $1.1202$ & $0.0029$ & $[1.0089,1.2222]$ & $0.0023$ & $0.0004$ & $0.0428$ \\
 & SMC ABC $\rho_e$             & $1.1221$ & $0.0030$ & $[1.0182,1.2361]$ & $0.0024$ & $0.0007$ & $0.0360$ \\
 & SMC ABC $\rho_H$             & $1.1199$ & $0.0029$ & $[1.0138,1.2228]$ & $0.0023$ & $0.0005$ & $0.0449$ \\
 & SMC ABC Regression $\rho_1$  & $1.1264$ & $0.0028$ & $[1.0165,1.2305]$ & $0.0022$ & $0.0001$ & $0.0198$ \\
 & SMC ABC Regression $\rho_e$  & $1.1287$ & $0.0029$ & $[1.0245,1.2383]$ & $0.0023$ & $0.0001$ & $0.0168$ \\
 & SMC ABC Regression $\rho_H$  & $1.1263$ & $0.0028$ & $[1.0299,1.2305]$ & $0.0022$ & $0.0001$ & $0.0218$ \\
\hline
\end{tabular} }}}
\end{table}

In order to highlight the importance of  the choice of the summary statistic in the output of ABC algorithms,  we also implemented them removing one coordinate in the proposed one, that is, we consider the summary statistics:
\begin{align*}
\mathcal{S}_1(\widetilde{\mathcal{Z}}_n)&=\left(\frac{\sum_{i=1}^n Z_i}{\sum_{i=0}^{n-1} Z_i},\frac{\phi_{n-1}(Z_{n-1})}{Z_{n-1}}\right),\\
\mathcal{S}_2(\widetilde{\mathcal{Z}}_n)&=\left(\sum_{i=1}^n Z_i,\frac{\phi_{n-1}(Z_{n-1})}{Z_{n-1}}\right),\\
\mathcal{S}_3(\widetilde{\mathcal{Z}}_n)&=\left(\sum_{i=1}^n Z_i,\frac{\sum_{i=1}^n Z_i}{\sum_{i=0}^{n-1} Z_i}\right).
\end{align*}
The results for the parameters $\theta$, $\gamma$ and $\tau_m$ are presented  in Tables \ref{tab:summary-theta}, \ref{tab:summary-gamma} and \ref{tab:summary-taum}, respectively.
   The tables   mainly reveal  that removing  the number of  progenitors in (at least) the  last generation from the observed sample and consequently from the summary statistic, that is $\mathcal{S}_3$,  provides similar  accuracy measures for $\tau_m$ -it is a consequence of the first convergence in (\ref{eq:conv-summary-stat}). However, the estimation of its factors separately, that is, $m$ (dependent on $\theta$) and $\gamma$ get worse. It is highlighted the less accurate estimate of the control parameter, $\gamma$, with a very wide 95\% HPD interval. If the component of total progeny is removed, that is $\mathcal{S}_1$, the ISE and KL for three parameters, $\theta$, $\gamma$, and $\tau_m$ increase significatively. For the latter parameter,  the 95\% HPD intervals do not contain the true value. Finally, removing the second coordinate in $\mathcal{S}$, that is $\mathcal{S}_2$, results in an increment of the three accuracy measures  for the three considered parameters.

%In the case of the summary statistic $\mathcal{S}_1(\cdot)$ one can observe an increment of the values of the RMSE, ISE and KL for the parameters $\theta$ and $\gamma$, but not for the parameter $\tau_m$. This indicates that these methods based on this summary statistic identify the parameter $\tau_m$ when only the population sizes are known, but they have more difficulty for the parameters $\theta$ and $\gamma$. This fact also shows the convenience of knowing the number of progenitors in the last generation. On the contrary, the use of the summary statistic $\mathcal{S}_2(\cdot)$ results in estimates of the posterior density with a smaller variance for the three parameters, which implies larger values of the ISE and KL and hence, the need of including the total progeny of the process in our summary statistic.}

To complete the study of this simulated example, we developed a sensitivity analysis on the choices of the prior distributions. Recall that both the offspring and control parameters are probabilities, and hence, beta distributions seem to be reasonable options  as prior distributions for both parameters. Thus, the aforementioned analysis was performed by considering different values for the shape parameters of beta distributions. The results of this analysis for the Gibbs sampler algorithm and the SMC ABC algorithm together with the local lineal regression adjustment using the metric $\rho_1$  are summarized in Tables \ref{tab:sens-theta-prior} and \ref{tab:sens-gamma-prior} for the posterior densities $\pi(\theta|\widetilde{\mathcal{Z}}_{30})$ and  $\pi(\gamma|\widetilde{\mathcal{Z}}_{30})$, respectively{, in  Supplementary Material}.
% since it is the ABC algorithm  that  provides the best estimate of the posterior densities according to Tables \ref{tab:summary-theta} and \ref{tab:summary-gamma}.
These results indicate that the estimation of the posterior densities {is} not sensitive to the choices of the prior distributions.

Finally, it is worth mentioning again that  ABC methodology relies on having ease of sampling from the model.  In our implementation, this implies the knowledge of the parametric families. An interesting issue is to study the performance of the algorithms when one knows that those distributions can be parametrized with a one-dimensional parameter, but the kind of parametric distribution one should use to that end is unknown. This issue is considered in the Supplementary Material.

\subsection{Real example: oligondrocyte cell population}
%quite accurate estimates

In the present example, we extend the algorithm proposed in a more complex model by considering a real data set belonging to the field of cell kinetics. The considered model  is a controlled two--type branching model. In addition to illustrate that the methodology can be extended without too much difficulty, the aim of this example is to illustrate  the ABC methodology  in a situation in which the true  posterior densities can be calculated. The population considered is a real population of oligodendrocyte cells already studied in \cite{art-Dposterior}, \cite{Hyrien-AMNY2006}, and \cite{Yakovlev-Stoimenova-Yanev-2008}. In these populations, one distinguishes two type of cells: the oligodendrocyte precursor cells -referred as $T_1$- and the terminally differentiated oligodendrocytes -referred as $T_2$. Regarding the reproduction, only the $T_1$ cells have reproductive capacity and they can produce two daughter cells of the same type, with probability $p_1$,  they can transform into a daughter cell of type $T_2$, with probability $p_2$, or they can even die without any offspring, with probability $p_0$. There is also an emigration component as a consequence of the migration of cells out of the field of observation. Let denote by $1-\gamma$ the probability of emigration of $T_1$-cells, therefore $\gamma$ represents the probability that a cell of type $T_1$ completes successfully its mitotic cycles regardless of its outcome. The parameters of interest are $\{p_0, p_1, p_2\}$ and $\gamma$.

With the aim of describing the proliferation of cells in these populations a continuous time branching process with emigration  and a controlled two-type process  for the embedded discrete structure were considered (see  \cite{art-Dposterior} and \cite{Yakovlev-Stoimenova-Yanev-2008} for further details). Briefly, focussing  on the embedded discrete structure, the controlled two-type branching process, denoted as $\{Z_n\}_{n\in\N_0}$, is defined as follows:
\vspace*{-1ex}
\begin{align}\label{eq:def-CMBP}
Z_0=(N_0,0),\quad Z_{n+1}= \sum_{j=1}^{\phi_n(Z_n)}(X_{n,j}^{(1)},X_{n,j}^{(2)}),\quad n\in\N_0,
\end{align}
with $Z_n=(Z_n^{(1)},Z_n^{(2)})$, $N_0\in\N$, and $\{(X_{n,j}^{(1)},X_{n,j}^{(2)}): j\in\N, n\in\N_0\}$ and $\{\phi_n(z): n\in\N_0, z\in\N_0^2\}$ being two independent families of non-negative integer valued random variables. Moreover, they are assumed to satisfy the next conditions:
\begin{enumerate}[label=(\roman*),ref=(\roman*),start=1]
\item For each $z=(z^{(1)},z^{(2)})\in\N_0^2$, the random variables $\{\phi_n(z): n\in\N_0\}$ are i.i.d. following a binomial distribution with parameters $z^{(1)}$ and $\gamma\in (0,1)$.

\item The stochastic processes $\{(X_{n,j}^{(1)},X_{n,j}^{(2)}): j\in\N\}$, $n\in\N_0$ are i.i.d. with probability distribution
\vspace*{-1ex}
\begin{align*}
p_0&=P\left[X_{n,j}^{(1)}=0,X_{n,j}^{(2)}=0\right],\\
p_1&=P\left[X_{n,j}^{(1)}=2,X_{n,j}^{(2)}=0\right],\\
p_2&=P\left[X_{n,j}^{(1)}=0,X_{n,j}^{(2)}=1\right].
\end{align*}
%$$p_0=P\left[X_{n,j}^{(1)}=0,X_{n,j}^{(2)}=0\right],\ p_1=P\left[X_{n,j}^{(1)}=2,X_{n,j}^{(2)}=0\right],\ p_2=P\left[X_{n,j}^{(1)}=0,X_{n,j}^{(2)}=1\right].$$
\vspace*{-2ex}
\item If $n_1,n_2\in\N_0$ are such that $n_1\neq n_2$, then, the sequences $\{(X_{n_1,j}^{(1)}, X_{n_1,j}^{(2)}): j\in\N\}$ and  $\{(X_{n_2,j}^{(1)},X_{n_2,j}^{(2)}): j\in\N\}$ are independent.
\end{enumerate}
 As we mentioned above, given the nature of the data, the sample constituted by the entire family tree is available, that is, the sample $\mathcal{Z}_n^*=\{Z_l^{(1)}(0),Z_l^{(1)}(2),\Lambda_l:\ l=0,\ldots,n-1\}$, with for $j=0,2$, and $l=0,\ldots, n-1$,
\begin{equation*}
Z_l^{(1)}(j)=\sum_{i=1}^{\phi_l(Z_l)} I_{\left\{X_{l,i}^{(1)}=j,\ X_{l,i}^{(2)}=0\right\}}\quad\mbox{ and }\quad  \Lambda_l=\sum_{i=1}^{\phi_l(Z_l)} I_{\left\{X^{(1)}_{l,i}=0,\ X^{(2)}_{l,i}=1\right\}}.
\end{equation*}
%and our aim is to estimate the posterior distribution of $(p_0,p_1,p_2,\gamma)$ upon on $\mathcal{Z}_n^*$, that is, $\pi\left(p_0,p_1,p_2,\gamma|\mathcal{Z}_n^*\right)$. To that end,
For $n\in\N$ and $j=0,2$, we introduce the variables:
$$Y_{n-1}^{(1)}(j)=\sum_{l=0}^{n-1} Z_l^{(1)}(j),\ \Psi_{n-1}=\sum_{l=0}^{n-1}\Lambda_l,\  \Delta_{n-1}=\sum_{l=0}^{n-1} \phi_l(Z_l),\ \text{ and }\  Y_{n-1}^{(1)}=\sum_{l=0}^{n-1} Z_l^{(1)},$$
where $Y_{n-1}^{(1)}(j)$ is the total number of cells of type $T_1$ producing exactly $j$ cells of type $T_1$, $j=0,2$, in the first $n-1$ generations, $\Psi_{n-1}$ is the total number of cells of type $T_1$ producing one cell of type $T_2$ in the first $n-1$ generations, $\Delta_{n-1}$ is the total number of progenitor cells of type $T_1$ in the first $n-1$ generations, and $Y_{n-1}^{(1)}$ is the total number of cells of type $T_1$ in the first $n-1$ generations. We propose as the  summary statistic
$$\mathcal{S}(\mathcal{Z}_n^*)=\left(Y_{n-1}^{(1)},\frac{Y_{n-1}^{(1)}(0)}{\Delta_{n-1}},\frac{Y_{n-1}^{(1)}(2)}{\Delta_{n-1}},\frac{\Delta_{n-1}}{Y_{n-1}^{(1)}}\right).$$
 This summary statistic is the equivalent in the present setting to the proposed summary statistic in Section \ref{sec:ABC}. The first component is the total progeny of the process in the first $n-1$ generations. The second and the third one are the relative proportion of cells of type $T_1$ having no offspring and two offspring of type $T_1$, respectively. The forth component is the proportion of cells of type $T_1$ that do not emigrate. Thus, we expect the second and third component to be informative on $p_0$ and $p_1$, respectively, and consequently on $p_2$, and the forth one to be helpful to identify $\gamma$. We also incorporate the first component to introduce the information about the total magnitude of the process. Notice that it is not required the incorporation of data from $T_2$  cells in the summary  statistic due their dependence with $T_1$ cells and the fact that these do not generate offspring.

The data sets that we study in this example resulted from one experiment started with 30 cells of type $T_1$ in a solution treated with a substance that boosts the production of $T_2$ cells, and whose family tree was observed until the generation $n=5$. These data are provided in Table \ref{tab:data-oligo} and were provided in \cite{Hyrien-AMNY2006}.

\begin{table}[H]
\centering
\caption{Data of the observed tree}\label{tab:data-oligo}
\begin{tabular}{|ccccccc|}
\cline{1-7}
 $n$ & $N$ & $Y_{n-1}^{(1)}$ & $\Delta_{n-1}$ & $Y_{n-1}^{(1)}(0)$ & $Y_{n-1}^{(1)}(2)$  & $\Psi_{n-1}$ \\
\hline
 $5$ & $30$ & $276$ & $269$ & $37$ & $133$ & $99$ \\
\hline
\end{tabular}
\end{table}

We ran the SMC ABC algorithm with the summary statistic proposed above and $M=3$ stages with a final local--regression adjustment. We considered  $\rho_1-$ distance in view of the comments of the simulated example. Thus, to obtain samples of size 2500 at each step, we generated pools of $10^5$, $10^6$, and $10^7$ non-extinct branching processes following the model in \eqref{eq:def-CMBP},  initiated with 30 cells of type $T_1$  and generated until the 5th generation and we took the quantiles of order $0.025$, $0.0025$, and $0.00025$  of the sample of the distances of the simulated processes as the thresholds $\epsilon_1$, $\epsilon_2$ and $\epsilon_3$ at each step,  respectively.  In particular, given the nature of our parameters we took a Dirichlet distribution with parameter $\alpha=(1/2,1/2,1/2)$ for the  offspring distribution $\{p_0,p_1,p_2\}$ and a beta distribution with parameters $1/2$ and $1/2$ as the prior distribution for $\gamma$. Note that in this case we can evaluate the accuracy of the estimates obtained using the ABC methodology by comparing them with the true ones obtained by using the theory of conjugate families. Indeed, by making use of Markov's property and the independence between the emigration and reproduction phases, one can prove that the likelihood function $f(\mathcal{Z}_n^*|p_0,p_1,\gamma)$ is given by
\begin{equation*}
f(\mathcal{Z}_n^*|p_0,p_1,\gamma)\propto p_{0}^{Y_{n-1}^{(1)}(0)} p_{1}^{Y_{n-1}^{(1)}(2)} (1-p_0-p_1)^{\Psi_{n-1}}\gamma^{\Delta_{n-1}} (1-\gamma)^{Y_{n-1}^{(1)}-\Delta_{n-1}},
\end{equation*}
and as a result, as indicated above, if we take as the prior distribution $\pi(p_0,p_1,\gamma)=\pi(p_0,p_1)\pi(\gamma)$, with
$$\pi(p_0,p_1)\propto p_0^{\alpha_0-1}p_1^{\alpha_1-1}(1-p_0-p_1)^{\alpha_2-1},\quad \text{ and }\quad\pi(\gamma)\propto \gamma^{\beta_1-1}(1-\gamma)^{\beta_2-1},$$
then, the posterior distribution is given by
\begin{eqnarray}
\pi(p_0,p_1,\gamma|\mathcal{Z}_n^*)&\propto& p_{0}^{Y_{n-1}^{(1)}(0)+\alpha_0-1} p_{1}^{Y_{n-1}^{(1)}(2)+\alpha_1-1} (1-p_0-p_1)^{\Psi_{n-1}+\alpha_2-1}\cdot\nonumber\\&& \cdot\gamma^{\Delta_{n-1}+\beta_1-1} (1-\gamma)^{Y_{n-1}^{(1)}-\Delta_{n-1}+\beta_2-1}\label{eq:post-conj}.
\end{eqnarray}
Figure \ref{real:fig-SMC-cont-exp} shows the contour plot of the offspring posterior density estimated by using the SMC ABC algorithm together local regression adjustment and one of the true density  provided by (\ref{eq:post-conj}). Moreover, it shows the estimated density and the true one of the parameter $\gamma$. The  estimated densities are centered around the parameters of interest with small dispersion, therefore the ABC methodology provides in this situation quite accurate estimates. This latter fact is also concluded by comparing the posterior means and posterior variances of $p_0$, $p_1$, $p_2$ and $\gamma$ obtained by the ABC methodology with the one obtained by using the theory of conjugate families (see Table \ref{real:tab-post-exp2}).

\begin{figure}[H]
\centering
\includegraphics[width=0.3\linewidth]{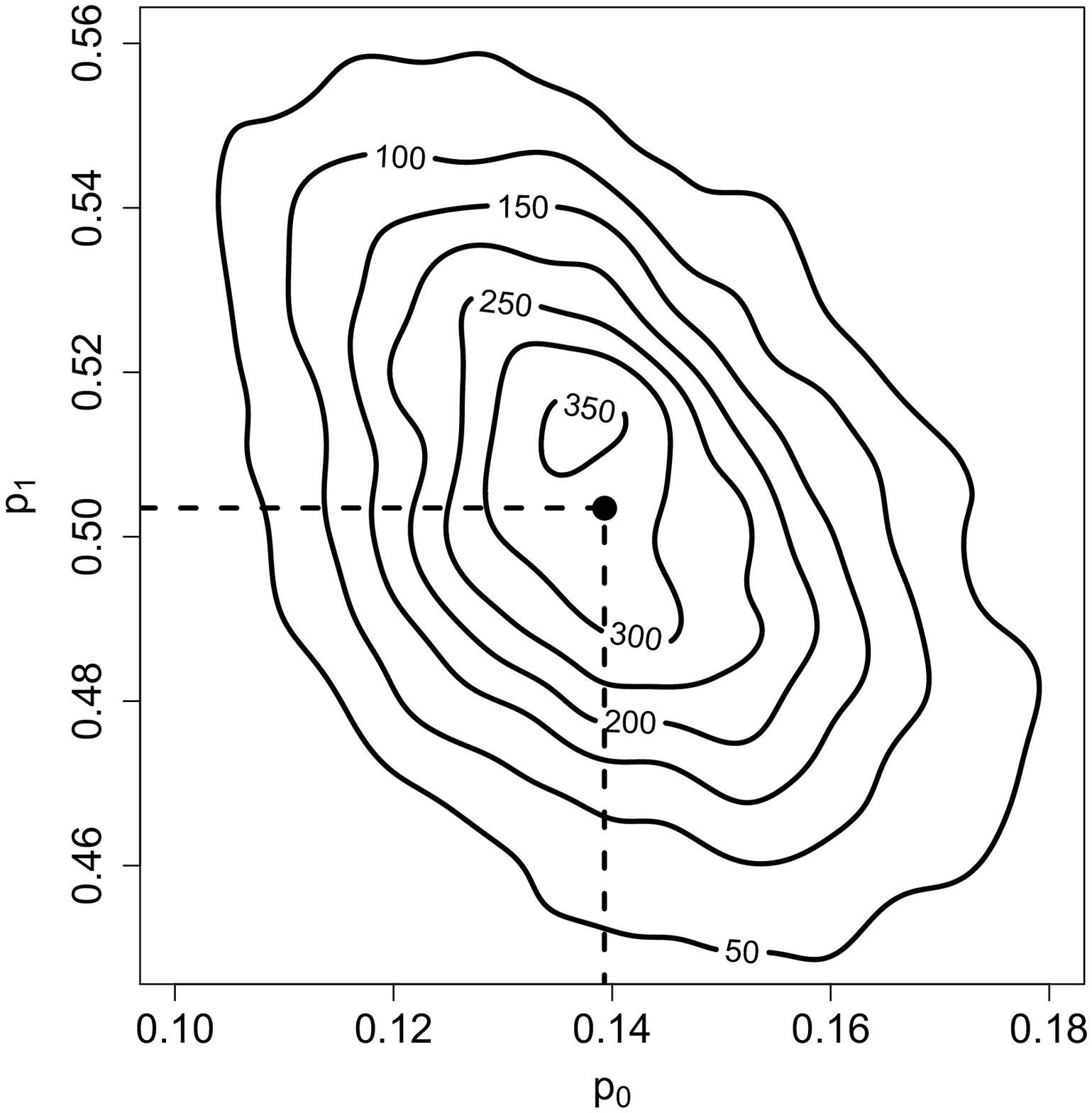}\hspace{0.5cm}
\includegraphics[width=0.3\linewidth]{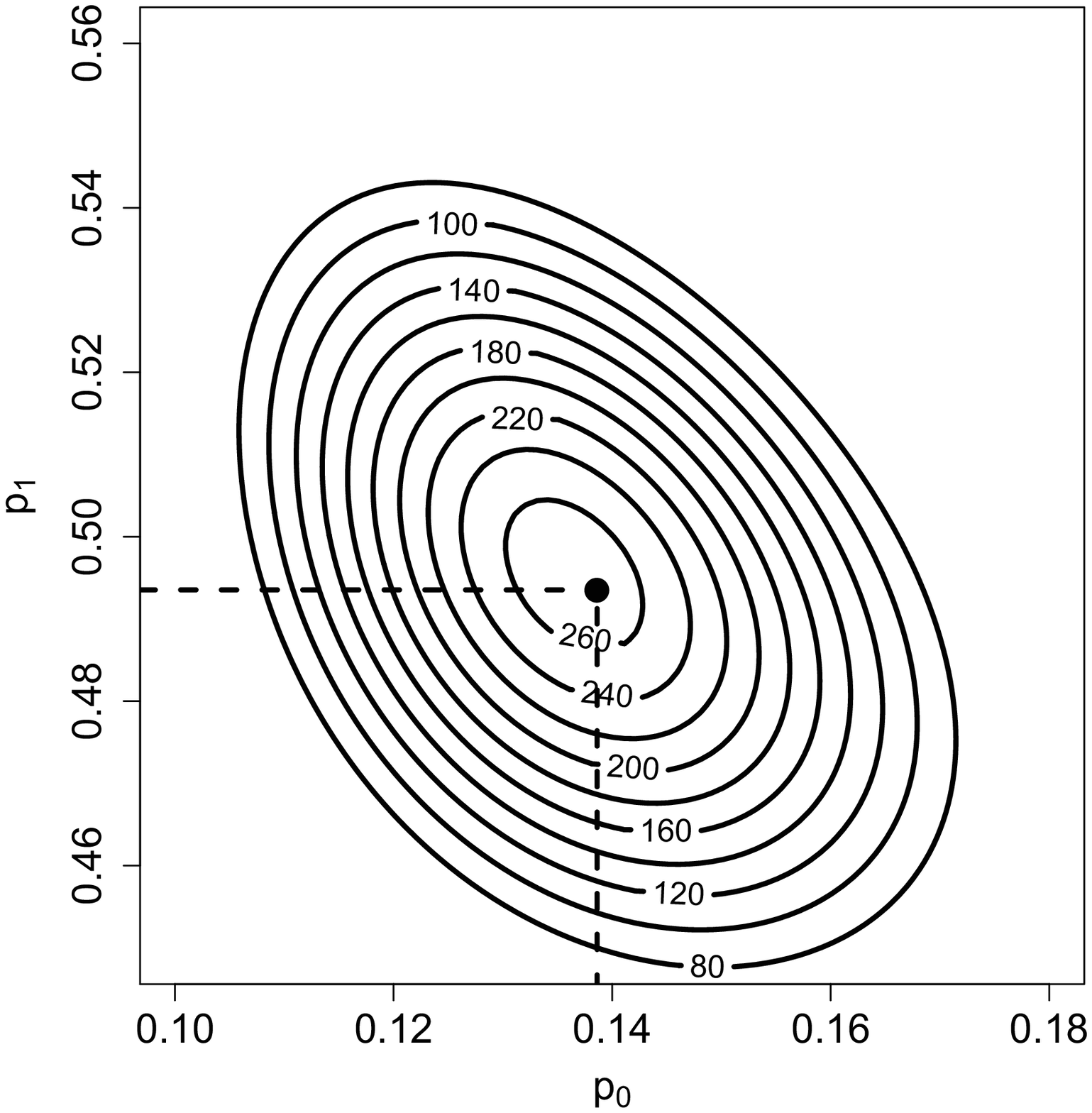}\hspace{0.5cm}
\includegraphics[width=0.3\linewidth]{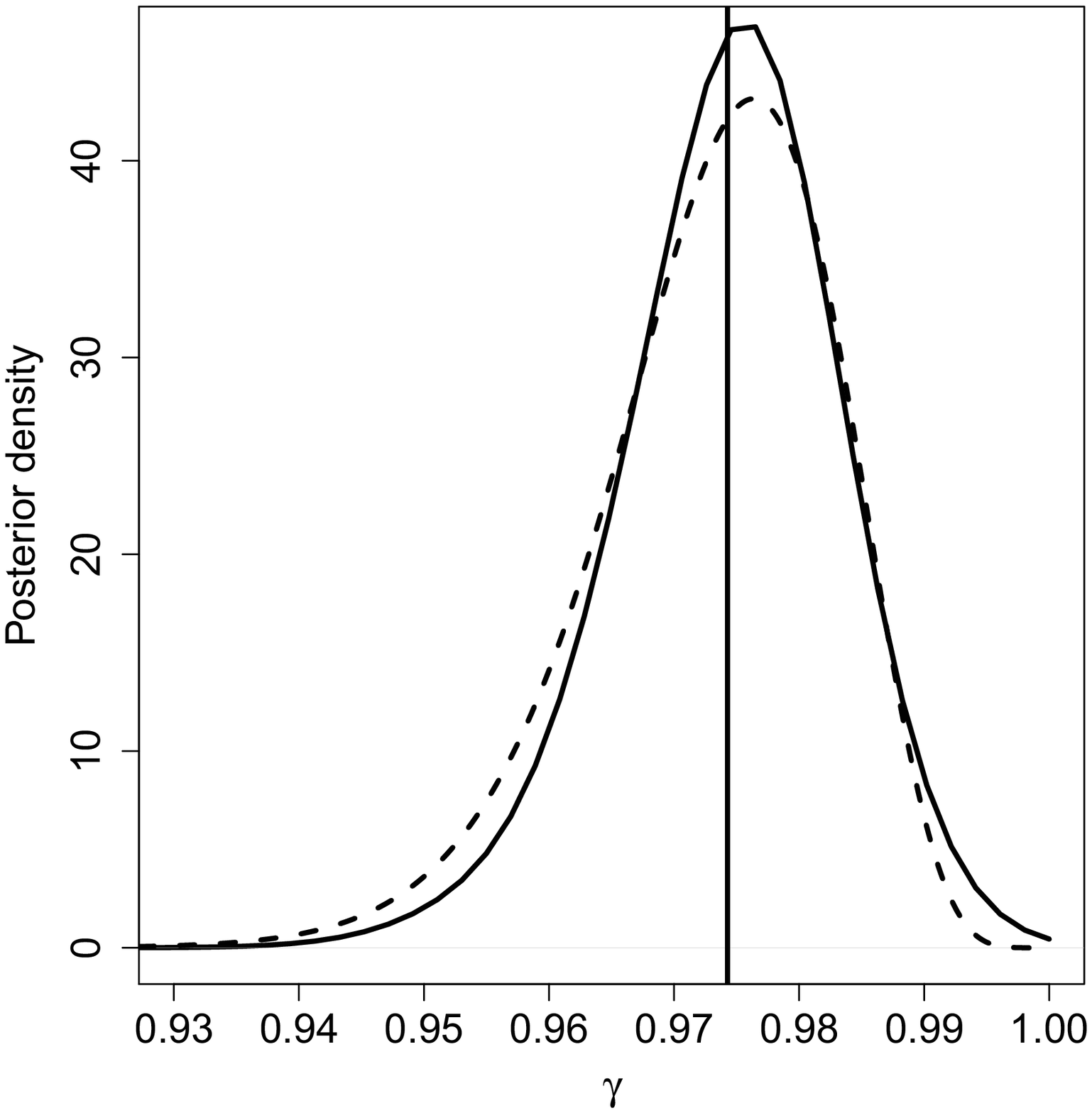}
\caption{Contour plot of the posterior density functions for the offspring distribution (solid lines) and posterior means (dotted lines) obtained by using the SMC ABC algorithm  together local regression adjustment (left) and the true posterior density (centre). Posterior density functions of $\gamma$ (solid line) and posterior mean (dotted line) obtained by using the SMC ABC algorithm together local regression adjustment and  dashed line represents the posterior density obtained using the theory of conjugate distributions.} \label{real:fig-SMC-cont-exp}
\end{figure}

\begin{table}[!htbp] \centering
  \caption{Comparison of the posterior means and variances}
  \label{real:tab-post-exp2}
\begin{tabular}{c|cccc|c|}

\cline{2-6}
& & $p_0$ & $p_1$ & $p_2$ & $\gamma$ \\
\hline
\multicolumn{1}{|c|}{\multirow{2}{*}{\parbox{2cm}{\centering SMC ABC algorithm}}}& \text{Mean} & $0.139296$ & $0.503495$ & $0.357209$ & $0.974284$   \\
\multicolumn{1}{|c|}{} &\text{Variance} & $0.000324$ & $0.000752$ & $0.000616$ & $0.000055$  \\
\hline
\multicolumn{1}{|c|}{\multirow{2}{*}{\parbox{2cm}{\centering True posterior}}} & \text{Mean}& $0.138697$ & $0.494012$ & $0.367291$ & $0.972987$  \\
\multicolumn{1}{|c|}{} & \text{Variance}& $0.000427$ & $0.000931$ & $0.000828$ & $0.000095$ \\
\hline
\end{tabular}
\end{table}

\begin{remark}\label{rem1}
The simulations were performed by using the statistical software \texttt{R} (see \cite{R}). For the convergence diagnostics of the Gibbs sampler algorithm we used the \texttt{coda} package  (see \cite{pbcv2010}). Regarding the multivariate normal distribution of algorithm SMC ABC, we made use of the function \texttt{dmvnorm()} of the \texttt{mvtnorm} package for the density function (see \cite{mvtnorm}) and the function \texttt{mvrnorm()} from the \texttt{MASS} package to draw pairs of numbers of such distributions (see \cite{mass}).
\end{remark}

\section{Concluding remarks}\label{sec:conclusions}

Motivated by the interest of making inference on the context of CBPs with random control functions, we have {explored} ABC methodology {and proposed an appropriate summary statistic for such a purpose in this setting}. Thus, this work constitutes a twofold generalization of \cite{art-ABC}: on the one hand, we considered a more complex model, CBPs with random control functions,
with the added difficulties that this implies, and on the other hand, we introduced {a suitable summary statistic to minimize the influence of the dimension of the data that usually arises when comparing data of large dimension}. To that end, we considered parametric frameworks for the offspring and control distributions and assumed that one can observe all the population sizes and the number of progenitors in (at least) the  last generation. It is worthwhile to highlight that the knowledge of the number of progenitors in the last generation, $\phi_{n-1}(Z_{n-1})$, and its inclusion in the summary statistic, plays an important role when identifying the true parameters of the model. This means a clear progress with respect to the aforesaid paper.

We have described an ABC algorithm to sample from the posterior distributions of the offspring and control parameters and  approximated the corresponding ones by making use of kernel density estimators. In particular, we have considered a SMC ABC algorithm followed by a local linear regression with a three-dimensional summary statistic. This methodology can be extended without too much difficulty to more complicated branching family processes, being  adequate to draw samples from posterior densities which are closer to the true posterior densities. Indeed, through an extensive analysis of a simulated example, the accuracy of the ABC procedure has been compared with the output of the Gibbs sampler.   We showed a visual comparison of the posterior density estimated by each algorithm and  a comparison based on summary statistics such as the posterior mean and variance, 95\% HPD intervals, RMSE, ISE and KL. The results of this example show that the SMC ABC algorithm with together with the local-linear regression post-processing   provides estimates of the posterior densities which are as accurate as the ones obtained with the Gibbs sampler and having the advantage of being computationally simpler. Moreover, a study of the influence of the choice of the distance, the prior distributions, and the offspring and control distribution considered in the model was carried out. It shows that this methodology is not sensible to such choices. Secondly, the adaptation of this methodology for a controlled two--type branching process is applied by considering a real data set, showing again that it suites quite well.
 We implemented these methodologies by using the statistical software and programming environment R.

\section*{Supplementary Material}

\section*{Simulated data}\label{subsec:sup:sim-data}

The data for the simulated example  in Subsection \ref{sec:example} are provided in Table \ref{tab:sim-data}. Recall that for the simulated CBP, which starts with $Z_0=1$ individual, the reproduction law  is a geometric distribution with parameter $q=0.4$, and for each $k\in\N_0$, the probability distribution of the control variable $\phi_n(k)$ is a binomial distribution with parameters $\xi(k)$ and $\gamma=0.75$, with $\xi(k)= k+\lfloor\log (k)\rfloor$, for each $k\in\N$ and $\xi(0)=0$.

\begin{table}[H]
\centering
\caption{Simulated data, $\widetilde{\mathcal{Z}}_{30}^{obs}$}
\label{tab:sim-data}
\begin{tabular}{|c|c|c|}
\hline
$n$ & $Z_n$ & $\phi_n(Z_n)$ \\
\hline
$0$ & $1$ & $1$ \\
$1$ & $4$ & $3$ \\
$2$ & $6$ & $5$ \\
$3$ & $4$ & $3$ \\
$4$ & $11$ & $10$ \\
$5$ & $6$ & $7$ \\
$6$ & $9$ & $7$ \\
$7$ & $19$ & $13$ \\
$8$ & $26$ & $19$ \\
$9$ & $14$ & $9$ \\
$10$ & $10$ & $9$ \\
$11$ & $11$ & $9$ \\
$12$ & $9$ & $7$ \\
$13$ & $12$ & $8$ \\
$14$ & $14$ & $12$ \\
$15$ & $15$ & $12$ \\
$16$ & $9$ & $5$ \\
$17$ & $3$ & $3$ \\
$18$ & $6$ & $7$ \\
$19$ & $13$ & $13$ \\
$20$ & $17$ & $15$ \\
$21$ & $23$ & $18$ \\
$22$ & $35$ & $32$ \\
$23$ & $58$ & $46$ \\
$24$ & $75$ & $61$ \\
$25$ & $73$ & $51$ \\
$26$ & $103$ & $78$ \\
$27$ & $107$ & $83$ \\
$28$ & $141$ & $100$ \\
$29$ & $166$ & $131$ \\
$30$ & $216$ & $\cdot$ \\
\hline
\end{tabular}
\end{table}

\section*{Sensitivity analysis}\label{subsec:sup:sensitivity}

A summary of the sensitivity analysis on the choice of the parameters of the prior distribution is presented in Tables \ref{tab:sens-theta-prior} and \ref{tab:sens-gamma-prior} for the posterior distribution of $\theta$ and $\gamma$, respectively. The results are provided for the Gibbs sampler algorithm and the SMC ABC algorithm  followed by the post-processing correction method using the metric $\rho_1$.

Finally, we present a summary of the results obtained from the sensitivity analysis on the choice of the parametric families for the offspring and control distributions for three different CBPs. Apart from the previous example, we simulated  30 generations of two additional CBPs starting with one individual: the first of them has a geometric distribution with parameter $\theta=0.918$ as the offspring distribution and the control variables $\phi_n(k)$ follow binomial distributions with parameters $\xi(k)$ and $\gamma=0.1$, and in the second CBP, the offspring distribution is a geometric distribution with parameter $\theta=0.556$ and it has control variables $\phi_n(k)$ following binomial distribution with parameters $\xi(k)$ and $\gamma=0.9$. We proposed two offspring laws (geometric and Poisson distributions) and three control laws (binomial, Poisson and negative binomial distribution) and ran the SMC ABC algorithm for the posterior densities $\pi(m|\widetilde{\mathcal{Z}}_{30})$, $\pi(\tau|\widetilde{\mathcal{Z}}_{30})$ and $\pi(\tau_m|\widetilde{\mathcal{Z}}_{30})$ in each one of these six situations for the three examples considered. We opted for these three parameters since it is more reasonable to compare the results based on different distributions in terms of the offspring mean, the parameter $\tau$ and the asymptotic mean growth rate of the process, which are the stable parameters of the process.

The results in Table \ref{tab:sens-model} show  that the method for the different models used for simulating usually identify the offspring mean, the parameter $\tau$ and the asymptotic mean growth rate relatively well, which also indicates the goodness of our summary statistic. A noteworthy result is the case of considering a Poisson distribution for the control laws. While the algorithm provides a good estimation for the posterior density of the parameter $\tau$, the posterior density for $m$ (and consequently, for $\tau_m$) does not seem to fit so well to the one obtained for the second simulated model when assuming a binomial or negative binomial distribution for the control laws. This is related to the magnitude of the parameter $\theta=0.918$. Indeed, when we used the geometric distribution for simulating the model we also estimated the posterior density of $\theta$, resulting a mean value of $0.9079$ and variance of $ 2.0629\cdot 10^{-05}$, which seems to be a good estimation of that posterior density. However, the small bias observed in this estimation was enlarged when we estimated $m$ due to the fact that $m=\theta(1-\theta)^{-1}$.

Similar results are obtained when one is unaware of the existence of the function $\xi(\cdot)$. To examine that problem in this example, we considered the three models described above and we repeated the study by applying same ABC method for the posterior distributions $\pi(m|\widetilde{\mathcal{Z}}_{30})$, $\pi(\tau|\widetilde{\mathcal{Z}}_{30})$ and $\pi(\tau_m|\widetilde{\mathcal{Z}}_{30})$. Again, we used geometric and Poisson offspring distributions and binomial, Poisson and negative binomial distributions as control laws. The results are presented in Table \ref{tab:sens-model-nog}, where one observes no significant difference with those in Table \ref{tab:sens-model}, except a slight increase in the RMSE in most of the cases. This similitude is caused by the fact that our model satisfies the condition \ref{rem:cond-convergence-a} in Remark \ref{rem:cond-convergence}.

\newpage

\begin{table}[H] \centering
\caption{Summary of the sensitivity analysis on the choice of the prior distribution for the posterior density $\pi(\theta|\widetilde{\mathcal{Z}}_{30})$}\label{tab:sens-theta-prior}
\scalebox{0.84}{{\renewcommand\baselinestretch{1.5}{\small\begin{tabular}{|c|c|c|c|c|c|c|}
\cline{2-7}
\multicolumn{1}{c|}{ } & $\pi(\theta)$ & $\pi(\gamma)$ & \multicolumn{4}{|c|}{$\pi(\theta|\widetilde{\mathcal{Z}}_{30})$}\\
\cline{2-7} \multicolumn{1}{c|}{ } & Beta distribution & Beta distribution & Mean & Variance & 95\% HPD & RMSE \\
\hline
\multirow{9}{*}{\rotatebox{90}{\centering Gibbs sampler}}
& $\boldsymbol{\beta}(0.5,3)$ & $\boldsymbol{\beta}(0.5,3)$        & $0.6007$ & $0.0002$ & $[0.5728,0.6275]$ & $0.0006$ \\
& $\boldsymbol{\beta}(0.5,3)$ & $\boldsymbol{\beta}(0.5,0.5)$      & $0.5996$ & $0.0002$ & $[0.5724,0.6283]$ & $0.0005$ \\
& $\boldsymbol{\beta}(0.5,3)$ & $\boldsymbol{\beta}(3,0.5)$        & $0.5993$ & $0.0002$ & $[0.5715,0.6257]$ & $0.0005$ \\
& $\boldsymbol{\beta}(0.5,0.5)$ & $\boldsymbol{\beta}(0.5,3)$      & $0.6035$ & $0.0002$ & $[0.5763,0.6305]$ & $0.0006$ \\
& $\boldsymbol{\beta}(0.5,0.5)$ & $\boldsymbol{\beta}(0.5,0.5)$    & $0.6007$ & $0.0002$ & $[0.5737,0.6275]$ & $0.0005$ \\
& $\boldsymbol{\beta}(0.5,0.5)$ & $\boldsymbol{\beta}(3,0.5)$      & $0.5997$ & $0.0002$ & $[0.5724,0.6270]$ & $0.0006$ \\
& $\boldsymbol{\beta}(3,0.5)$ & $\boldsymbol{\beta}(0.5,3)$        & $0.6040$ & $0.0002$ & $[0.5780,0.6318]$ & $0.0006$ \\
& $\boldsymbol{\beta}(3,0.5)$ & $\boldsymbol{\beta}(0.5,0.5)$      & $0.6016$ & $0.0002$ & $[0.5737,0.6279]$ & $0.0005$ \\
& $\boldsymbol{\beta}(3,0.5)$ & $\boldsymbol{\beta}(3,0.5)$        & $0.6008$ & $0.0002$ & $[0.5732,0.6292]$ & $0.0006$ \\
\hline
\multirow{9}{*}{\rotatebox{90}{\centering SMC ABC Regression $\rho_1$}}
& $\boldsymbol{\beta}(0.5,3)$ & $\boldsymbol{\beta}(0.5,3)$        & $0.5956$ & $0.0002$ & $[0.5691,0.6212]$ & $0.0005$ \\
& $\boldsymbol{\beta}(0.5,3)$ & $\boldsymbol{\beta}(0.5,0.5)$      & $0.5952$ & $0.0002$ & $[0.5696,0.6212]$ & $0.0006$ \\
& $\boldsymbol{\beta}(0.5,3)$ & $\boldsymbol{\beta}(3,0.5)$        & $0.5938$ & $0.0002$ & $[0.5687,0.6194]$ & $0.0006$ \\
& $\boldsymbol{\beta}(0.5,0.5)$ & $\boldsymbol{\beta}(1,3)$        & $0.5976$ & $0.0002$ & $[0.5718,0.6247]$ & $0.0005$ \\
& $\boldsymbol{\beta}(0.5,0.5)$ & $\boldsymbol{\beta}(0.5,0.5)$    & $0.5958$ & $0.0002$ & $[0.5705,0.6225]$ & $0.0005$ \\
& $\boldsymbol{\beta}(0.5,0.5)$ & $\boldsymbol{\beta}(3,0.5)$      & $0.5963$ & $0.0002$ & $[0.5705,0.6229]$ & $0.0005$ \\
& $\boldsymbol{\beta}(3,0.5)$ & $\boldsymbol{\beta}(0.5,3)$        & $0.5979$ & $0.0002$ & $[0.5714,0.6261]$ & $0.0005$ \\
& $\boldsymbol{\beta}(3,0.5)$ & $\boldsymbol{\beta}(0.5,0.5)$      & $0.5962$ & $0.0002$ & $[0.5705,0.6234]$ & $0.0005$ \\
& $\boldsymbol{\beta}(3,0.5)$ & $\boldsymbol{\beta}(3,0.5)$        & $0.5956$ & $0.0002$ & $[0.5691,0.6203]$ & $0.0005$ \\
\hline
\end{tabular}}}}
\end{table}

\begin{table}[H]
\caption{Summary of the sensitivity analysis on the choice of the prior distribution for the posterior density $\pi(\gamma|\widetilde{\mathcal{Z}}_{30})$}\label{tab:sens-gamma-prior}
\centering
\scalebox{0.84}{{\renewcommand\baselinestretch{1.5}{\small\begin{tabular}{|c|c|c|c|c|c|c|}
\cline{2-7}
\multicolumn{1}{c|}{ } & $\pi(\theta)$ & $\pi(\gamma)$ & \multicolumn{4}{|c|}{$\pi(\gamma|\widetilde{\mathcal{Z}}_{30})$}\\
\cline{2-7}
\multicolumn{1}{c|}{ } & Beta distribution & Beta distribution & Mean & Variance & 95\% HPD & RMSE \\
\hline
\multirow{9}{*}{\rotatebox{90}{\centering Gibbs sampler}}
 & $\boldsymbol{\beta}(0.5,3)$ & $\boldsymbol{\beta}(0.5,3)$        & $0.7526$ & $0.0010$ & $[0.6913,0.8142]$ & $0.0018$ \\
 & $\boldsymbol{\beta}(0.5,3)$ & $\boldsymbol{\beta}(0.5,0.5)$      & $0.7560$ & $0.0010$ & $[0.6961,0.8181]$ & $0.0018$ \\
 & $\boldsymbol{\beta}(0.5,3)$ & $\boldsymbol{\beta}(3,0.5)$        & $0.7582$ & $0.0009$ & $[0.6996,0.8172]$ & $0.0017$ \\
 & $\boldsymbol{\beta}(0.5,0.5)$ & $\boldsymbol{\beta}(0.5,3)$      & $0.7442$ & $0.0009$ & $[0.6834,0.8032]$ & $0.0017$ \\
 & $\boldsymbol{\beta}(0.5,0.5)$ & $\boldsymbol{\beta}(0.5,0.5)$    & $0.7548$ & $0.0009$ & $[0.6957,0.8133]$ & $0.0017$ \\
 & $\boldsymbol{\beta}(0.5,0.5)$ & $\boldsymbol{\beta}(3,0.5)$      & $0.7577$ & $0.0010$ & $[0.6974,0.8199]$ & $0.0018$ \\
 & $\boldsymbol{\beta}(3,0.5)$ & $\boldsymbol{\beta}(0.5,3)$        & $0.7431$ & $0.0009$ & $[0.6856,0.8028]$ & $0.0017$ \\
 & $\boldsymbol{\beta}(3,0.5)$ & $\boldsymbol{\beta}(0.5,0.5)$      & $0.7532$ & $0.0010$ & $[0.6909,0.8146]$ & $0.0018$ \\
 & $\boldsymbol{\beta}(3,0.5)$ & $\boldsymbol{\beta}(3,0.5)$        & $0.7563$ & $0.0010$ & $[0.6957,0.8177]$ & $0.0018$ \\
 \hline
\multirow{9}{*}{\rotatebox{90}{\centering SMC ABC Regression $\rho_1$}}
 & $\boldsymbol{\beta}(0.5,3)$ & $\boldsymbol{\beta}(0.5,3)$        & $0.7592$ & $0.0009$ & $[0.6994,0.8213]$ & $0.0018$ \\
 & $\boldsymbol{\beta}(0.5,3)$ & $\boldsymbol{\beta}(0.5,0.5)$      & $0.7618$ & $0.0010$ & $[0.7025,0.8213]$ & $0.0019$ \\
 & $\boldsymbol{\beta}(0.5,3)$ & $\boldsymbol{\beta}(3,0.5)$        & $0.7656$ & $0.0009$ & $[0.7039,0.8239]$ & $0.0021$ \\
 & $\boldsymbol{\beta}(0.5,0.5)$ & $\boldsymbol{\beta}(0.5,3)$      & $0.7542$ & $0.0011$ & $[0.6896,0.8177]$ & $0.0019$ \\
 & $\boldsymbol{\beta}(0.5,0.5)$ & $\boldsymbol{\beta}(0.5,0.5)$    & $0.7613$ & $0.0010$ & $[0.6990,0.8253]$ & $0.0020$ \\
 & $\boldsymbol{\beta}(0.5,0.5)$ & $\boldsymbol{\beta}(3,0.5)$      & $0.7596$ & $0.0009$ & $[0.6999,0.8181]$ & $0.0018$ \\
 & $\boldsymbol{\beta}(3,0.5)$ & $\boldsymbol{\beta}(0.5,3)$        & $0.7545$ & $0.0010$ & $[0.6888,0.8181]$ & $0.0019$ \\
 & $\boldsymbol{\beta}(3,0.5)$ & $\boldsymbol{\beta}(0.5,0.5)$      & $0.7615$ & $0.0010$ & $[0.7008,0.8217]$ & $0.0020$ \\
 & $\boldsymbol{\beta}(3,0.5)$ & $\boldsymbol{\beta}(3,0.5)$        & $0.7627$ & $0.0010$ & $[0.7016,0.8239]$ & $0.0020$ \\
\hline
\end{tabular}}}}
\end{table}

\begin{table}[H]
\caption{Summary of the sensitivity analysis on the choice of the offspring and control distributions  in the SMC ABC algorithm with the summary statistic and the post-processing method. A geometric distribution and a Poisson distribution were fitted for the reproduction law, while a binomial distribution, a Poisson distribution and a negative binomial distribution were considered for the control laws}\label{tab:sens-model}
\centering
\scalebox{0.75}{{\renewcommand\baselinestretch{1.5}{\small\begin{tabular} {|c|c|c|c|c|c|c|c|c|}
\cline{4-9}
%\multicolumn{3}{c|}{ } & \multicolumn{6}{|c|}{Fitted distributions}\\ \cline{4-9}
\multicolumn{3}{c|}{ } & \multicolumn{3}{|c|}{Geometric offspring distribution} & \multicolumn{3}{|c|}{Poisson offspring distribution}\\
\hline
Model & Parameter & Summary & $B(\xi(k),\rho)$ & $\mathcal{P}(\xi(k)\lambda)$ & $NB(\xi(k),\varrho)$ & $B(\xi(k),\rho)$ & $\mathcal{P}(\xi(k)\lambda)$ & $NB(\xi(k),\varrho)$ \\
\hline
\multirow{9}{*}{\rotatebox{90}{\parbox{4cm}{\centering $X_{ni}\sim G(0.6)$,  $\phi_n(k)\sim B(\xi(k),0.75)$}}}  & \multirow{3}{*}{{\centering $m=1.5$}}
  & Mean      & $1.4814$ & $1.4627$ & $1.5429$ & $1.4973$ & $1.5151$ & $1.5867$ \\
& & Variance  & $0.0067$ & $0.0190$ & $0.0342$ & $0.0044$ & $0.0127$ & $0.0303$ \\
& & RMSE      & $0.0031$ & $0.0091$ & $0.0160$ & $0.0020$ & $0.0057$ & $0.0168$ \\
\cline{2-9}
& \multirow{3}{*}{{\centering $\tau=0.75 $}}
  & Mean     & $0.7577$ & $0.7627$ & $0.7297$ & $0.7591$ & $0.7472$ & $0.7170$ \\
& & Variance & $0.0010$ & $0.0051$ & $0.0072$ & $0.0010$ & $0.0035$ & $0.0065$ \\
& & RMSE     & $0.0018$ & $0.0093$ & $0.0134$ & $0.0019$ & $0.0063$ & $0.0135$ \\
\cline{2-9}
& \multirow{3}{*}{{\centering $\tau_m=1.125$}}
  & Mean     & $1.1209$ & $1.1073$ & $1.1123$ & $1.1351$ & $1.1264$ & $1.1251$ \\
& & Variance & $0.0023$ & $0.0035$ & $0.0041$ & $0.0013$ & $0.0022$ & $0.0031$ \\
& & RMSE     & $0.0019$ & $0.0030$ & $0.0033$ & $0.0011$ & $0.0017$ & $0.0025$ \\
\hline
\multirow{9}{*}{\rotatebox{90}{\parbox{4cm}{\centering $X_{ni}\sim G(0.918)$, $\phi_n(k)\sim B(\xi(k),0.1)$}}} & \multirow{3}{*}{{\centering $m=11.25$}}
  & Mean     & $11.1465$ & $9.9684$ & $11.1538$ & $11.1516$ & $7.4936$ & $10.9934$ \\
& & Variance & $0.2964$ & $0.2929$ & $0.3813$ & $0.2546$ & $0.0171$ & $0.3343$ \\
& & RMSE     & $0.0024$ & $0.0143$ & $0.0031$ & $0.0020$ & $0.1095$ & $0.0030$ \\
\cline{2-9}
& \multirow{3}{*}{{\centering $\tau=0.1$}}
  & Mean     & $0.0979$ & $0.0987$ & $0.0981$ & $0.0981$ & $0.1033$ & $0.0996$ \\
& & Variance & $0.00002$ & $0.0001$ & $0.00003$ & $0.00002$ & $0.00005$ & $0.00003$ \\
& & RMSE     & $0.0028$ & $0.0052$ & $0.0032$ & $0.0025$ & $0.0059$ & $0.0028$ \\
\cline{2-9}
& \multirow{3}{*}{{\centering $\tau_m=1.125$}}
  & Mean     & $1.0885$ & $0.9813$ & $1.0912$ & $1.0919$ & $0.7734$ & $1.0926$ \\
& & Variance & $0.0006$ & $0.0017$ & $0.0007$ & $0.0004$ & $0.0017$ & $0.0004$ \\
& & RMSE     & $0.0013$ & $0.0166$ & $0.0012$ & $0.0009$ & $0.0969$ & $0.0009$ \\
\hline
\multirow{9}{*}{\rotatebox{90}{\parbox{4cm}{\centering $X_{ni}\sim G(0.556)$,  $\phi_n(k)\sim B(\xi(k),0.9)$}}} & \multirow{3}{*}{{\centering $m=1.25$}}
  & Mean     & $1.2574$ & $1.2511$ & $1.2562$ & $1.2571$ & $1.2570$ & $1.2628$ \\
& & Variance & $0.0006$ & $0.0021$ & $0.0033$ & $0.0004$ & $0.0017$ & $0.0033$ \\
& & RMSE     & $0.0004$ & $0.0013$ & $0.0021$ & $0.0002$ & $0.0011$ & $0.0022$ \\
\cline{2-9}
& \multirow{3}{*}{{\centering $\tau=0.9$}}
  & Mean     & $0.9098$ & $0.9136$ & $0.9098$ & $0.9097$ & $0.9100$ & $0.9065$ \\
& & Variance & $0.0001$ & $0.0010$ & $0.0017$ & $0.0001$ & $0.0009$ & $0.0018$ \\
& & RMSE     & $0.0002$ & $0.0015$ & $0.0022$ & $0.0002$ & $0.0012$ & $0.0022$ \\
\cline{2-9}
& \multirow{3}{*}{{\centering $\tau_m=1.125$}}
  & Mean     & $1.1439$ & $1.1419$ & $1.1409$ & $1.1434$ & $1.1428$ & $1.1425$ \\
& & Variance & $0.0004$ & $0.0006$ & $0.0007$ & $0.0002$ & $0.0004$ & $0.0006$ \\
& & RMSE     & $0.0005$ & $0.0006$ & $0.0007$ & $0.0004$ & $0.0005$ & $0.0006$ \\
\hline
\end{tabular}}}}
\end{table}

\begin{table}[H]
\caption{Summary of the sensitivity analysis on the choice of the offspring and control distributions  in the SMC ABC algorithm with the summary statistic and the post-processing method without the knowledge of the function $\xi(\cdot)$. A geometric distribution and a Poisson distribution were fitted for the reproduction law, while a binomial distribution, a Poisson distribution and a negative binomial distribution were considered for the control laws}\label{tab:sens-model-nog}
\centering
\scalebox{0.82}{{\renewcommand\baselinestretch{1.5}{\small\begin{tabular} {|c|c|c|c|c|c|c|c|c|}
\cline{4-9}
%\multicolumn{3}{c|}{ } & \multicolumn{6}{|c|}{Fitted distributions}\\ \cline{4-9}
\multicolumn{3}{c|}{ } & \multicolumn{3}{|c|}{Geometric offspring distribution} & \multicolumn{3}{|c|}{Poisson offspring distribution}\\
\hline
Model & Parameter & Summary & $B(k,\rho)$ & $\mathcal{P}(k\lambda)$ & $NB(k,\varrho)$ & $B(k,\rho)$ & $\mathcal{P}(k\lambda)$ & $NB(k,\varrho)$ \\
\hline
\multirow{9}{*}{\rotatebox{90}{\parbox{4cm}{\centering $X_{ni}\sim G(0.6)$,  $\phi_n(k)\sim B(\xi(k),0.75)$}}}  & \multirow{3}{*}{{\centering $m=1.5$}}
& Mean           & $1.5315$ & $1.4884$ & $1.5659$ & $1.5439$ & $1.5428$ & $1.6178$ \\
& & Variance     & $0.0067$ & $0.0183$ & $0.0336$ & $0.0047$ & $0.0132$ & $0.0302$ \\
& & RMSE         & $0.0034$ & $0.0082$ & $0.0169$ & $0.0029$ & $0.0067$ & $0.0196$ \\
\cline{2-9}
& \multirow{3}{*}{{\centering $\tau=0.75 $}} & Mean  & $1.2054$ & $1.1926$ & $1.1942$ & $1.2109$ & $1.2069$ & $1.2072$ \\
& & Variance                                         & $0.0027$ & $0.0041$ & $0.0050$ & $0.0014$ & $0.0027$ & $0.0038$ \\
& & RMSE                                             & $0.0073$ & $0.0069$ & $0.0077$ & $0.0069$ & $0.0074$ & $0.0084$ \\
\cline{2-9}
& \multirow{3}{*}{{\centering $\tau_m=1.125$}} & Mean  & $1.2064$ & $1.1912$ & $1.1928$ & $1.2121$ & $1.2111$ & $1.2081$ \\
& & Variance                                           & $0.0027$ & $0.0040$ & $0.0048$ & $0.0014$ & $0.0026$ & $0.0038$ \\
& & RMSE                                               & $0.0073$ & $0.0066$ & $0.0074$ & $0.0071$ & $0.0079$ & $0.0085$ \\
\hline
\multirow{9}{*}{\rotatebox{90}{\parbox{4cm}{\centering $X_{ni}\sim G(0.918)$, $\phi_n(k)\sim B(\xi(k),0.1)$}}} & \multirow{3}{*}{{\centering $m=11.25$}}
& Mean                                           & $11.1775$ & $9.7894$ & $11.2174$ & $11.1970$ & $7.0385$ & $11.0671$ \\
& & Variance                                     & $0.3018$ & $0.2564$ & $0.3879$ & $0.2373$ & $0.0138$ & $0.3241$ \\
& & RMSE                                         & $0.0024$ & $0.0178$ & $0.0031$ & $0.0019$ & $0.1380$ & $0.0027$ \\
\cline{2-9}
& \multirow{3}{*}{{\centering $\tau=0.1$}} & Mean  & $0.0981$ & $0.1007$ & $0.0979$ & $0.0980$ & $0.1110$ & $0.0995$ \\
& & Variance                                       & $0.00002$ & $0.00005$ & $0.00003$ & $0.00002$ & $0.00005$ & $0.00003$ \\
& & RMSE                                           & $0.0027$ & $0.0048$ & $0.0033$ & $0.0023$ & $0.0170$ & $0.0027$ \\
\cline{2-9}
& \multirow{3}{*}{{\centering $\tau_m=1.125$}} & Mean  & $1.0943$ & $0.9833$ & $1.0955$ & $1.0949$ & $0.7809$ & $1.0980$ \\
& & Variance                                           & $0.0007$ & $0.0018$ & $0.0007$ & $0.0004$ & $0.0014$ & $0.0004$ \\
& & RMSE                                               & $0.0011$ & $0.0162$ & $0.0010$ & $0.0008$ & $0.0926$ & $0.0007$ \\
\hline
\multirow{9}{*}{\rotatebox{90}{\parbox{4cm}{\centering $X_{ni}\sim G(0.556)$,  $\phi_n(k)\sim B(\xi(k),0.9)$}}} & \multirow{3}{*}{{\centering $m=1.25$}}
& Mean                                                  & $1.2723$ & $1.2657$ & $1.2692$ & $1.2724$ & $1.2672$ & $1.2737$ \\
& & Variance                                            & $0.0006$ & $0.0021$ & $0.0035$ & $0.0004$ & $0.0018$ & $0.0035$ \\
& & RMSE                                                & $0.0006$ & $0.0015$ & $0.0024$ & $0.0005$ & $0.0013$ & $0.0025$ \\
\cline{2-9}
& \multirow{3}{*}{{\centering $\tau=0.9$}} & Mean  & $0.9159$ & $0.9211$ & $0.9195$ & $0.9166$ & $0.9221$ & $0.9168$ \\
& & Variance                                       & $0.0001$ & $0.0011$ & $0.0018$ & $0.0001$ & $0.0010$ & $0.0020$ \\
& & RMSE                                           & $0.0004$ & $0.0019$ & $0.0027$ & $0.0004$ & $0.0018$ & $0.0028$ \\
\cline{2-9}
& \multirow{3}{*}{{\centering $\tau_m=1.125$}} & Mean  & $1.1652$ & $1.1646$ & $1.1649$ & $1.1662$ & $1.1674$ & $1.1654$ \\
& & Variance                                           & $0.0004$ & $0.0006$ & $0.0008$ & $0.0002$ & $0.0004$ & $0.0006$ \\
& & RMSE                                               & $0.0014$ & $0.0016$ & $0.0018$ & $0.0014$ & $0.0016$ & $0.0017$ \\
\hline
\end{tabular}}}}
\end{table}

\end{document}